\documentclass{article}

\usepackage{arxiv}

\usepackage[utf8]{inputenc} % allow utf-8 input
\usepackage[T1]{fontenc}    % use 8-bit T1 fonts
\usepackage{hyperref}       % hyperlinks
\usepackage{url}            % simple URL typesetting
\usepackage{booktabs}       % professional-quality tables
\usepackage{amsfonts}       % blackboard math symbols
\usepackage{nicefrac}       % compact symbols for 1/2, etc.
\usepackage{microtype}      % microtypography
\usepackage{lipsum}		% Can be removed after putting your text content
\usepackage{graphicx}
\usepackage{natbib}
\usepackage{doi}

\usepackage{amsmath}
\usepackage{graphicx}
\usepackage{gensymb}
\usepackage{caption}
\usepackage{subcaption}
\usepackage{natbib}
\usepackage{nameref}

\graphicspath{ {images/} }

\title{Projecting U.S. coastal storm surge risks and impacts with deep learning}

\author{
    Julian R. Rice\thanks{Corresponding author: \texttt{julian.rice@pnnl.gov}}
    \hspace{3em}
    Karthik Balaguru
    \hspace{3em}
    Fadia {Ticona Rollano}
    \hspace{3em}
    John Wilson\\
    \textbf{Brent Daniel
    \hspace{3em}
    David Judi
    \hspace{3em}
    Ning Sun
    \hspace{3em}
    L. Ruby Leung}\\
    Pacific Northwest National Laboratory\\
    902 Battelle Boulevard, Richland, WA\\
}

\begin{document}
\maketitle 

\begin{abstract}
Storm surge is one of the deadliest hazards posed by tropical cyclones (TCs), yet assessing its current and future risk is difficult due to the phenomenon's rarity and physical complexity. Recent advances in artificial intelligence applications to natural hazard modeling suggest a new avenue for addressing this problem. We utilize a deep learning storm surge model to efficiently estimate coastal surge risk in the United States from 900,000 synthetic TC events, accounting for projected changes in TC behavior and sea levels. The derived historical 100-year surge (the event with a 1\% yearly exceedance probability) agrees well with historical observations and other modeling techniques. When coupled with an inundation model, we find that heightened TC intensities and sea levels by the end of the century result in a 50\% increase in population at risk. Key findings include markedly heightened risk in Florida, and critical thresholds identified in Georgia and South Carolina.
\end{abstract}

\section{Introduction}

% setting up the problem: TCs and storm surge are deadly, and rare
Tropical cyclones (TCs, also known as hurricanes or typhoons) have been the most damaging single form of weather-related natural disaster globally in terms of both loss of life and economic damage in recent decades \citep{world_meteorological_organization_wmo_wmo_2021}. In the United States (U.S.), storm surge has been the leading cause of deaths directly attributable to TCs, making up nearly as many deaths as all other TC-induced hazards combined \citep{rappaport_fatalities_2014}. With the growing density of people and assets in coastal regions \citep{klotzbach_continental_2018}, paired with increasing TC intensities in the future \citep{knutson_tropical_2020}, the threat posed by TC-induced storm surge may only grow more pronounced. However, deadly storm surge is such a rare occurrence that the historical record is not an accurate measure of the true risk---and certainly does not capture the potential for future changes in TC behavior. While flood risk is most often communicated in terms of the 100-year (1\% yearly exceedance probability) flood, such as in the Federal Emergency Management Agency (FEMA) Flood Insurance Rate Maps, the record of reliable surge observations in the U.S. is at most 150 years (generously, 1880--present \citep{needham_data-driven_2014}). Analyzing the 100-year event as a Bernoulli process suggests that the true 100-year return level has never been met or exceeded in this historical window for 22\% of the U.S. coastline.

% physical-numerical models are good, but slow
% introduce deep learning as alternative
Given insufficient observational records, the other avenue to understanding future storm surge risk is via modeling. While simple parametric models of storm surge exist \citep{irish_influence_2008, islam_new_2021}, the physical complexity of the processes involved render them incapable of sufficiently accurate spatial modeling. Physical-numerical models such as the Advanced Circulation (ADCIRC) model \citep{luettich_adcirc_1992, pringle_global_2021}, Delft3D \citep{roelvink_design_1995}, the Finite-Volume Coastal Ocean Model (FVCOM) \citep{chen_unstructured_2003}, and the Regional Ocean Model (ROMS) \citep{shchepetkin_regional_2005} are capable of highly accurate storm surge (and in some cases inland inundation) modeling. Yet, this accuracy requires a tradeoff with computational efficiency; due to the complex physics involved, these models often simulate fluid dynamics on the timescale of seconds on detailed meshes with many thousands of nodes. The computational expense associated with simulating thousands of years of TC activity are thus immense, and usually preclude a Monte Carlo-style probabilistic uncertainty analysis. Only in the last handful of years have data-driven models (namely, deep neural networks) presented a viable alternative to numerical modeling approaches, as illustrated by recent advances in neural networks for global weather forecasting \citep{pathak_fourcastnet_2022, bi_accurate_2023, lam_learning_2023, price_gencast_2024} and general circulation modeling \citep{kochkov_neural_2024}, often with orders-of-magnitude speedups in prediction time. Theoretical analysis of these data-driven models indicates that sufficient training and model complexity allows them to implicitly encode the underlying physics of these systems \citep{rupe_nonequilibrium_2022}.

% method overview
Previous deep learning approaches to storm surge modeling have shown promise, though these past methods have all been spatially inflexible either by training a model for only a single bay or subregion \citep{sztobryn_forecast_2003, lee_neural_2006, oliveira_neural_2009, hashemi_efficient_2016, sahoo_prediction_2019, lee_rapid_2021, xie_developing_2023, adeli_advanced_2023} or by training separate models for each location of interest \citep{lockwood_using_2022} thereby foregoing the benefits of learning shared physics. In contrast, we propose a ``point-based'' approach, which enables our model to be highly flexible, parallelizable, and capable of learning more complete and generalized physics of storm surge across all locations. Our proposed model, which we call DeepSurge, is a recurrent neural network trained to predict the peak surge level at any given location in the North Atlantic basin, trained and validated on ADCIRC outputs from more than 250 historical TC storm surge events. Although by no means a perfect surrogate for ADCIRC, DeepSurge achieves reasonably skillful out-of-sample accuracy (81.5\% $R^2$, 0.25~m mean absolute error) and up to a 96x speedup in prediction time compared with ADCIRC. Further, DeepSurge and ADCIRC show comparable skill when validated against independent National Oceanic and Atmospheric Administration (NOAA) tide gauge observations. 

% introduce RAFT tracks
For a robust quantification of risk, many thousands of TCs are required, necessitating the use of synthetic TC events. For this task, we use the Risk Analysis Framework for Tropical Cyclones (RAFT) \citep{xu_north_2024} to generate 900,000 TCs ($\sim$60,000 simulation years) representative of historical and future conditions, to our knowledge an order of magnitude larger than any previous synthetic TC-driven storm surge risk assessment. From these storms, we use DeepSurge to robustly estimate surge heights along the entire U.S. Gulf and Atlantic coastline, and combine with probabilistic sea-level rise projections and an efficient inland inundation model to characterize the extremes of storm surge flooding risk impacts.

\section{Methods}

\subsection{Numerical storm surge simulations}
A large and varied dataset of historical storm surge data was generated with the ADCIRC model \citep{luettich_jr_solution_1991, luettich_adcirc_1992, pringle_global_2021}, utilizing a mesh of 15,467 nodes spanning the North Atlantic basin. Historical TC data for the North Atlantic was retrieved from the International Best Track Archive for Climate Stewardship (IBTrACS) \citep{knapp_international_2010, knapp_international_2018}. Wind forcings were generated with the method described in \citet{emanuel_self-stratification_2011}, and pressure forcing with the methodology presented by \citet{holland_revised_2008}. Simulations were run at 1-second time steps for the lifetime of each storm, with the maximum water level at each node retained as the training target for our deep learning surrogate model. Further details are provided in Supplementary Section 1.

\subsection{DeepSurge, a deep learning storm surge model}
We develop a neural network model informed by the physics of the problem, which we call DeepSurge. Because storm surge is the result of complex physical interactions in space (coastal geometry and bathymetry) and time (storm evolution), our model utilizes both convolutional and recurrent neural network components which are known to be skilled in processing these forms of data respectively \citep{oshea_introduction_2015, hochreiter_long_1997} and have been successfully utilized in previous storm surge modeling approaches \citep{adeli_advanced_2023, xie_developing_2023, giaremis_storm_2024}. Our model differs from others in that it operates in a ``point-based'' manner, meaning it processes each node independently from its neighbors, which allows more flexibility, generalizability, and parallelization than previous approaches. The DeepSurge model ingests a timeseries and spatial maps describing a single node during a storm, and predicts that node's maximum surge level.

% NN architecture
\begin{figure}[t]
    \centering
    \includegraphics[width=\textwidth]{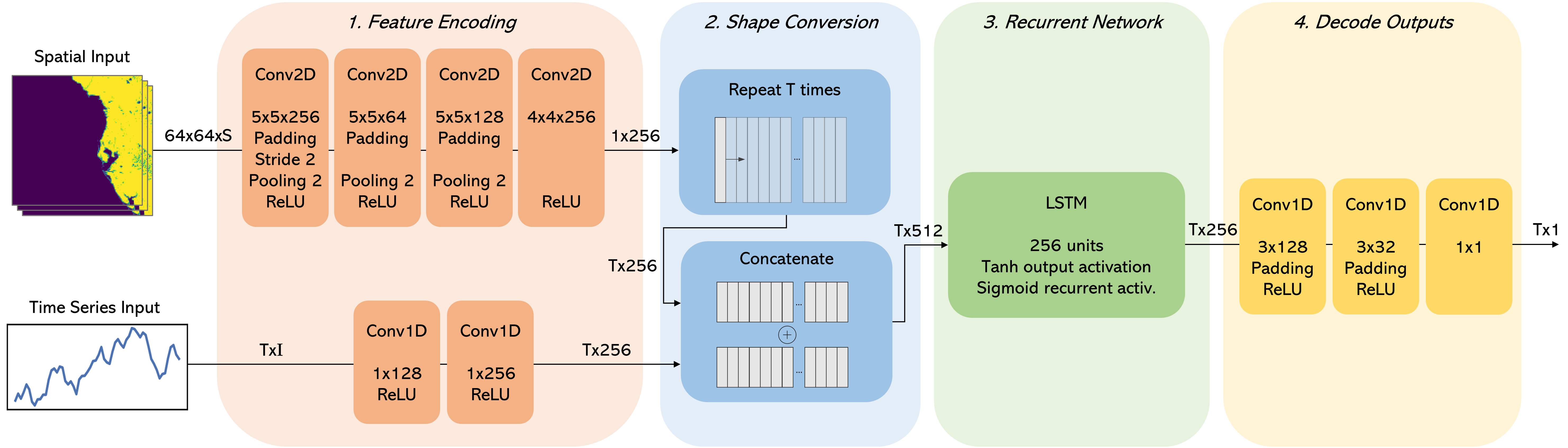}
    \caption{DeepSurge architecture. Tensor shapes (batch size not included) are given for each arrow, representing that tensor being passed from one layer to the next. In the tensor shapes, $S$ is the number of spatial images, $T$ is the number of timesteps, and $I$ is the number of timeseries inputs. Convolutions are summarized with kernel size $K$ and number of filters $F$, e.g. $K$x$K$x$F$ for a 2D convolution}
    \label{fig:neural-net-arch}
\end{figure}

The basic structure of the neural network, as detailed in Figure \ref{fig:neural-net-arch}, has four stages: 1) encode the timeseries and spatial data separately, 2) convert them to compatible shapes and concatenate them together, 3) apply a Long Short-Term Memory (LSTM) \citep{hochreiter_long_1997} layer to understand the temporal development, and 4) decode the features to predict a maximum surge level. The timeseries input to the model consists of a wide variety of variables describing the time evolution of the storm and its relation to the node of interest: storm maximum wind speed, storm radius of maximum wind, distance from node of interest to storm center, direction from node of interest to storm center, storm translation speed, and direction of storm motion. Also included are a few time-invariant scalars: bathymetric depth at the node of interest and estimated slope and direction of the seafloor. All directions and angles are encoded as sine/cosine pairs to avoid discontinuities. The spatial inputs to the model consist of a log-scaled bathymetric map interpolated from the ADCIRC mesh, and a land-ocean mask derived from 300-m resolution European Space Agency global land cover data \citep{european_space_agency_land_2017}. Both are 128x128 pixel grids at $0.0222\degree$ resolution centered on the node of interest, resulting in roughly a 300~km diameter receptive field. The network is constructed and trained in TensorFlow/Keras \citep{martin_abadi_tensorflow_2015}, and has a total of 1.7 million parameters. Details of the architecture and training methodology are provided in Supplementary Section 2.2.

\subsection{Validation \& sensitivity analysis of deep learning method}
DeepSurge demonstrates promising generalization skill on the test set, which consists of 71 storms that the model has never been trained or tuned on. When compared to the corresponding ADCIRC simulations, DeepSurge achieves an 81.5\% $R^2$ score, a mean squared error of 0.224~m, and a mean absolute error of 0.258~m in predicting peak surge heights per node. We additionally perform a validation against NOAA tide gauge observations\footnote{Accessible from the Tides \& Currents API: \url{https://api.tidesandcurrents.noaa.gov/mdapi/prod/\#}}. Although tide gauges are quite accurate under normal conditions, they often fail or malfunction during extreme surge events, which causes weaker events to be sampled more reliably than stronger ones. Overall, DeepSurge shows reasonable skill in capturing these gauge-observed surge peaks, with error metrics ($R^2=0.403$, $\text{MAE}=0.474$m, $\text{RMSE}=0.664$m) similar to the ADCIRC simulations ($R^2=0.427$, $\text{MAE}=0.543$m, $\text{RMSE}=0.882$m), and highly significant Pearson and Spearman correlations ($p \ll 0.001$). Possibly due to the negative sampling bias of the gauge observations, both DeepSurge and ADCIRC exhibit similar positive mean biases ($+0.217$m and $+0.203$m respectively). Sensitivity analysis suggests this bias may affect estimated inundation totals on the order of 14-18\%. See Supplementary Section 2.4 for further details of the tide gauge comparison and sensitivity analysis.

\subsection{Quantifying storm surge risk with synthetic tropical cyclones}
Synthetic tropical cyclones are generated with the the Risk Analysis Framework for Tropical Cyclones (RAFT) \citep{xu_deep_2021, balaguru_increased_2023, xu_north_2024}, forced by the climate conditions from nine Coupled Model Intercomparison Project Phase 6 (CMIP6) general circulation models (GCMs). The RAFT synthetic TC method produces realistic and diverse synthetic TC geneses, tracks, and intensities \citep{xu_north_2024}, and has been used in past studies to assess changes in TC landfall frequency \citep{balaguru_increased_2023}, wind turbine damage \citep{lipari_amplified_2024}, and power outage risk \citep{rice_projected_2025}. RAFT is forced with the climate conditions from a historical (1980--2014) and end-of-century future (2066--2100) period under Shared Socioeconomic Pathway SSP5-8.5 from nine CMIP6 models. We generate 50,000 TCs from the forcings of each of the 18 CMIP6 model-time period pairs, for a total of 900,000 tracks. Quantile delta mapping \citep{cannon_bias_2015} is applied to the TC intensities to correct for biases in the CMIP6 forcings. The set of synthetic TCs used in this study are identical to those described in and utilized by \citet{lipari_amplified_2024} and \citet{rice_projected_2025}. For further details on these synthetic storms, see those publications and Supplementary Section 1.

For each of these storms, DeepSurge is used to predict peak storm surge levels at 1,100 points along the U.S. Gulf and Atlantic coast. Projected surge levels in the future period are combined with the probabilistic sea-level rise projections from \citet{kopp_probabilistic_2014} to create a joint probability distribution at every coastal location. As in other studies \citep{gori_tropical_2022}, surge and sea-level rise are treated as linearly additive, and the future distributions of TC climatology and sea-level rise as independent (which may result in generally conservative estimates of future change \citep{little_joint_2015}).

\subsection{CA-Surge, a simple inland inundation model}
Lastly, we evaluate the impacts of modeled extreme water levels with a simple and efficient inundation model, CA-Surge, which puts projected changes in terms of the number of people impacted and enables analysis of sensitivity to various sources of uncertainty. CA-Surge is a ``bathtub-style" model similar to previously published approaches \citep{strauss_tidally_2012, yunus_uncertainties_2016, williams_comparative_2020} with the addition of an overland attenuation factor, which accounts for the reduction of surge height as water moves inland due to bottom friction. Attenuation rates are gathered from \citet{vafeidis_water-level_2019}, though it should be noted that there is substantial uncertainty inherent in attenuation rate estimates. Inundated areas are combined with LandScan \citep{dobson_landscan_2000} night-time population maps for estimates of affected population (for interpretability, we hold population constant between the historical and future periods). This method is found to produce reasonable estimates of inundated area compared with observations from Hurricane Katrina and captures state-level patterns in populations at risk from the 100-year coastal flood as derived by FEMA (see Supplementary Section 3.2). Still, we stress that this is an approximate method, useful primarily for efficient assessment of large-scale surge inundation. Pseudo-code of the CA-Surge algorithm is provided in Supplementary Section 3.1.

\section{Results}

The combination of RAFT, DeepSurge, and CA-Surge makes tractable the efficient simulation of many thousands of storm events, which enables robust estimation of the effect of changes in TC behavior on once-in-a-century coastal flood extremes. In this section, we evaluate the historical 100-year surge heights, analyze how they are projected to change in the future, and then put these changes in terms of population at risk of flooding.

\subsection{Extreme storm surge heights}

\begin{figure}
    \centering

    \begin{subfigure}[b]{0.8\textwidth}
        \centering
        \includegraphics[width=\textwidth]{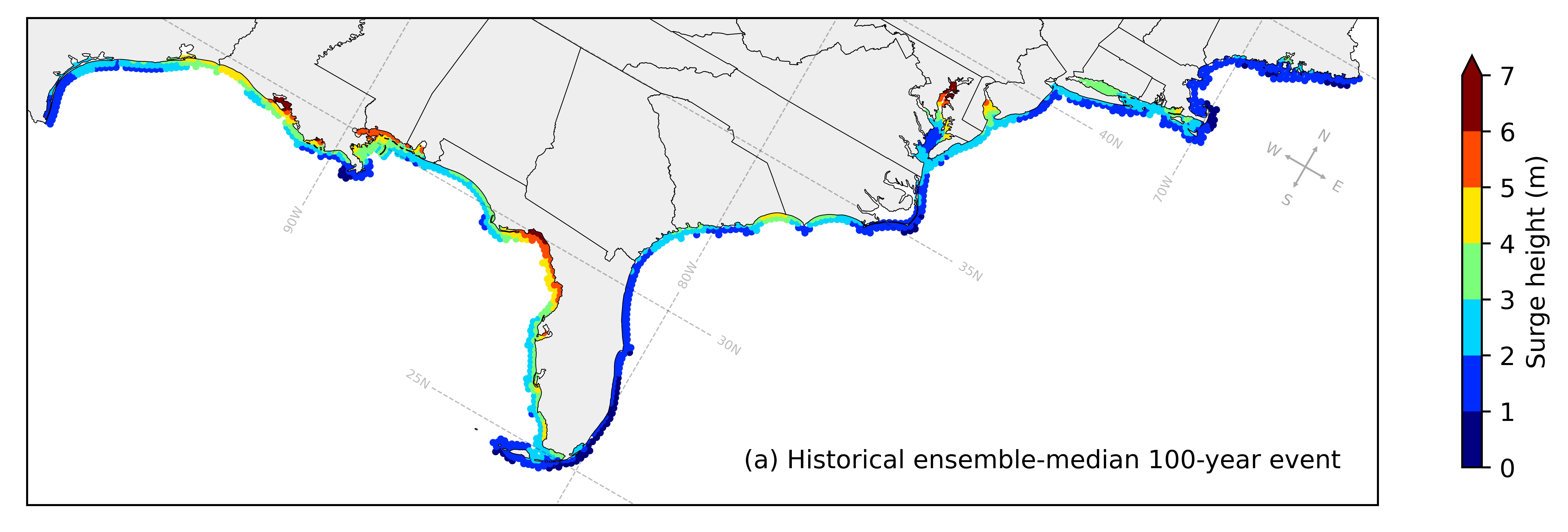}
        \phantomsubcaption
        % \caption{Historical}
        \label{fig:deepsurge-hist}
    \end{subfigure}

    \begin{subfigure}[b]{0.49\textwidth}
        \centering
        \includegraphics[width=\textwidth]{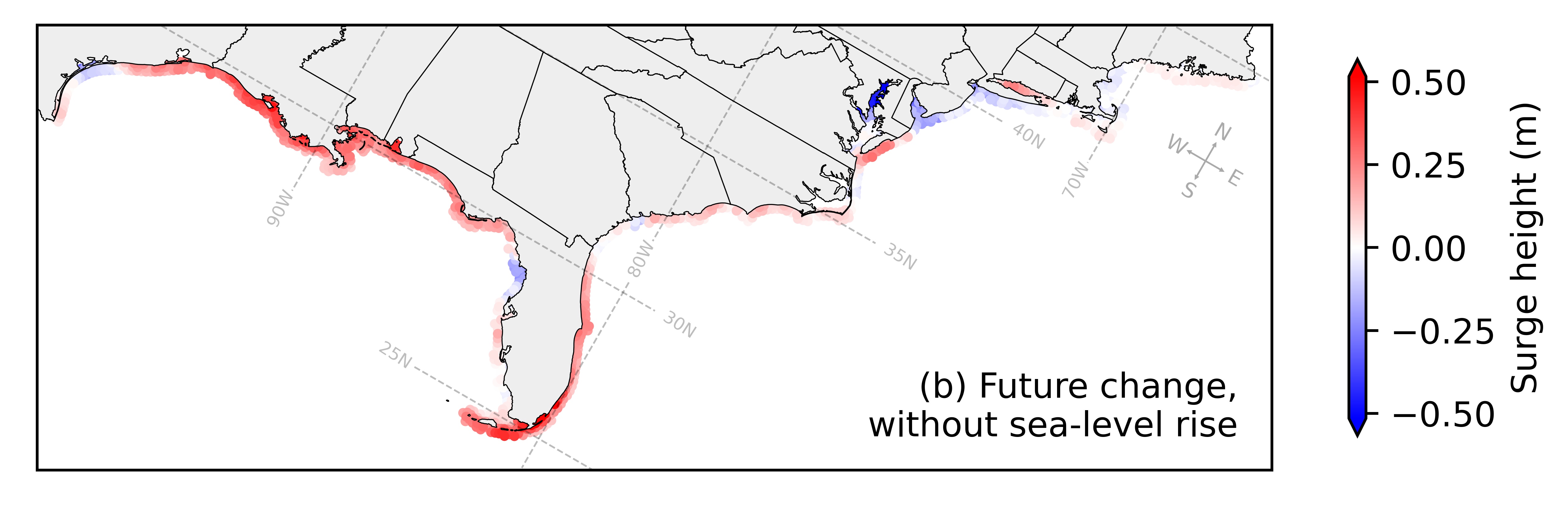}
        \phantomsubcaption
        % \caption{Future change, without sea-level rise}
        \label{fig:deepsurge-futu-change-noslr}
    \end{subfigure}
    \hfill
    \begin{subfigure}[b]{0.49\textwidth}
        \centering
        \includegraphics[width=\textwidth]{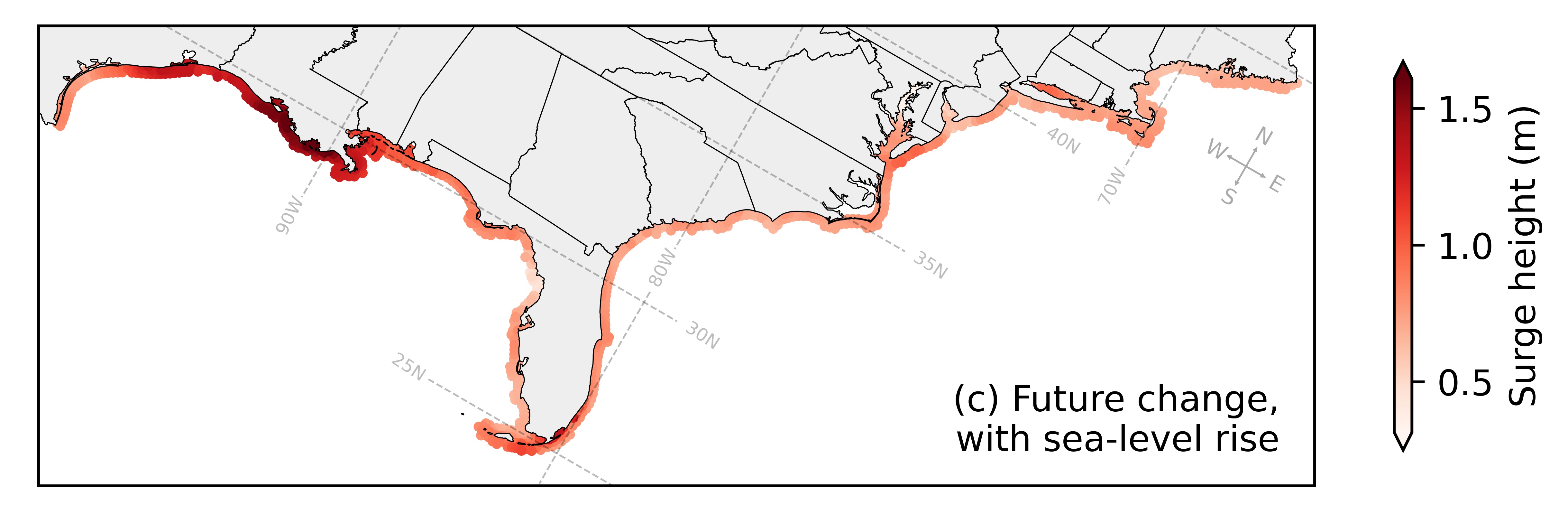}
        \phantomsubcaption
        % \caption{Future change, with sea-level rise}
        \label{fig:deepsurge-futu-change-slr}
    \end{subfigure}

    \begin{subfigure}[b]{0.49\textwidth}
        \centering
        \includegraphics[width=\textwidth]{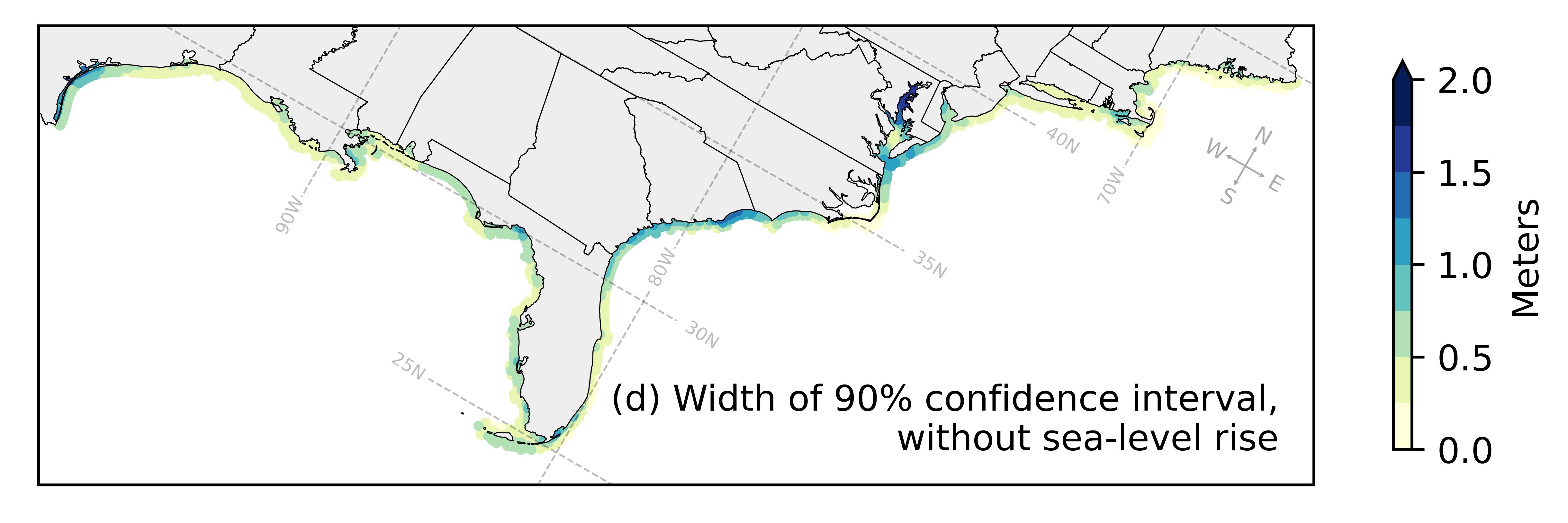}
        \phantomsubcaption
        % \caption{Width of 90\% confidence interval, without sea-level rise}
        \label{fig:deepsurge-futu-change-noslr-ci}
    \end{subfigure}
    \hfill
    \begin{subfigure}[b]{0.49\textwidth}
        \centering
        \includegraphics[width=\textwidth]{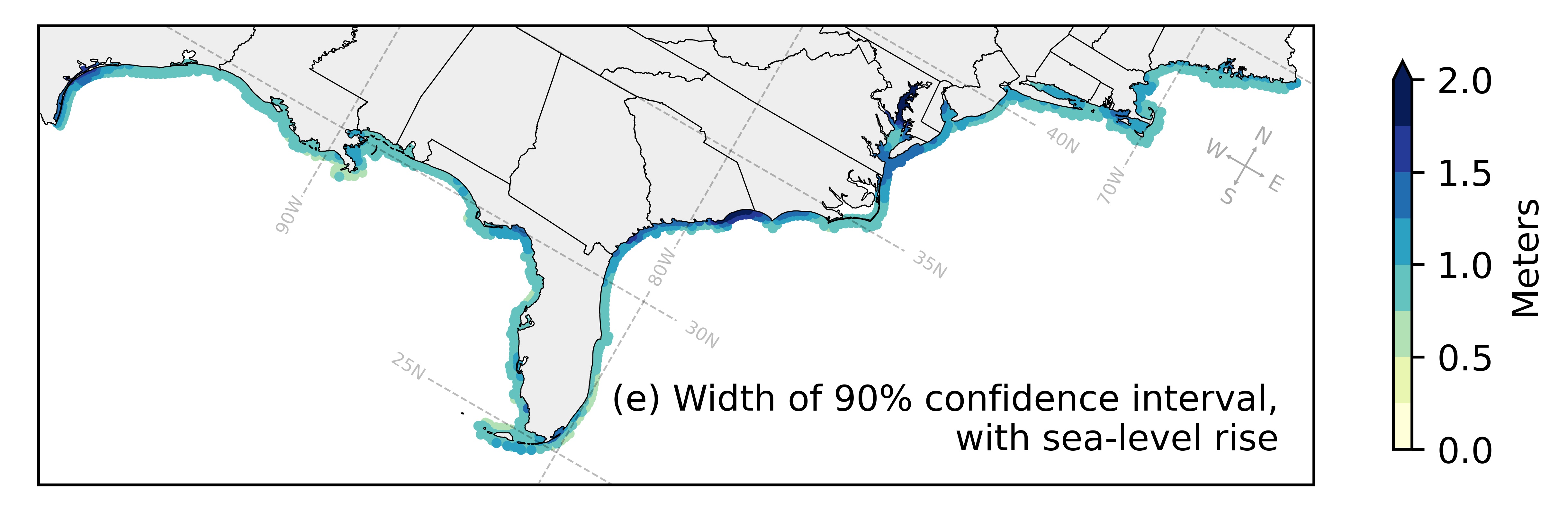}
        % \caption{Width of 90\% confidence interval, with sea-level rise}
        \phantomsubcaption
        \label{fig:deepsurge-futu-change-slr-ci}
    \end{subfigure}
     
    \caption{(\textbf{a}) DeepSurge modeled historical ensemble-median 100-year event; the corresponding future change (\textbf{b}) with and (\textbf{c}) without sea-level rise; and respective widths of the 90\% confidence intervals (\textbf{d},\textbf{e}).}
    \label{fig:deepsurge-rp100}
\end{figure}

% historical 100-year event
The ensemble median 100-year surge event as modeled by our method for the historical period (Fig. \ref{fig:deepsurge-hist}) shows predictable patterns. The highest surges occur primarily in the Gulf Coast where landfalling major hurricanes are most common, with peaks clustered in bays and concave coastlines. From the east coast of Florida and northward along the Atlantic, the combination of a steeper continental shelf and fewer major hurricanes results in lower extremes, with notable exceptions in areas with more complex coastal geometries including the Chesapeake Bay, Delaware Bay, and Long Island Sound. However, the results in the Chesapeake Bay appear to be anomalous; we find that this outlier is inherited from the original ADCIRC simulations, which had insufficient spatial resolution to resolve the fine-scale riverine and estuarine hydrodynamic processes in this region. Addressing this issue is planned for future work. Results in the Maryland-Virginia region are provided in this remainder of this work for completeness but should be treated with caution due to these biases. Otherwise, the spatial pattern of our modeled 100-year event appears qualitatively reasonable.

% other 100-year event estimations
To quantitatively verify that the combination of DeepSurge and RAFT are producing accurate 100-year surge heights, we compare our ensemble median return levels with the historical 100-year event observational estimates from \citet{needham_data-driven_2014}, who undertook an exhaustive analysis of the historical storm surge record 1900--2014 throughout the Gulf of Mexico, and three of independent surge modeling techniques: \citet{gori_tropical_2022} applied ADCIRC to a similar synthetic TC model forced with NCEP reanalysis; \citet{muis_global_2023} forced the Global Tide and Surge Model (a hydrodynamic model based on Delft3D) with historical ERA5 reanlysis; and lastly, for a comparison with a simple parametric model we generate surges with the Storm Surge Hazard Potential Index (SSHPI) developed by \citet{islam_new_2021}, forced by the same RAFT synthetic TCs as DeepSurge. We find that our method matches the observed distribution of 100-year surge heights at least as well as the other modeling methods, and shows strong and statistically significant correlation with the patterns produced by them (Spearman correlations $\ge0.5$, $p < 0.05$; see Supplementary Section 2.5 for details). DeepSurge achieves this comparable level of skill while being much more computationally efficient than numerical hydrodynamic models (up to 96x faster than our ADCIRC configuration; see Supplementary Section 2.3).

% future changes
In the future period, RAFT projects broadly increasing coastal TC intensities, with the 100-year storm intensity increasing in strength by roughly one Saffir-Simpson category in most coastal regions (Supplementary Fig. 1). The synthetic TCs also move slower on average in the future period, with slight differences in storm movement direction most visible in the vicinity of Florida (Supplementary Fig. 2). These differences in TC behavior cause substantial changes in storm surge as modeled by DeepSurge. Even without sea-level rise, the model indicates that these differences in TC behavior will produce notably larger 100-year surge levels across the northern Gulf Coast and eastern Florida (up to $+78$~cm), with more heterogeneous results for the rest of the coastline (Fig. \ref{fig:deepsurge-futu-change-noslr}). These results are in broad agreement with past studies such as \citet{gori_tropical_2022}. Causal analysis suggests that the increase in future TC intensities is the dominant factor contributing to these changes, while decreasing storm translation speeds are generally a weakly negative factor (Supplementary Section 4). The westward shift in average storm direction in the vicinity of Florida (Supplementary Fig. 2) may explain the differing responses on the peninsula's eastern and western coasts despite similar increases in TC intensities. Overall, altered TC behavior is projected to be a positive factor, increasing surge height by an average of $+8.4$~cm. 

The inclusion of sea-level rise (Fig. \ref{fig:deepsurge-futu-change-slr}) results in much larger changes, with a mean increase of $+85$~cm and maximum of $+170$~cm relative to historical, though with correspondingly larger uncertainty (Fig. \ref{fig:deepsurge-futu-change-slr-ci}). While the effect of sea-level rise is generally larger than that of future TC behavior (roughly two times larger at the peak along the Louisiana coast), the latter is still a significant contributor especially in capturing differences across finer spatial scales.

\subsection{Changing coastal inundation risk}
% introduce CA-Surge
While the quantification of future changes in extreme surge heights is of vital importance, it lacks crucial context; due to widely varying topographies and population densities, the relationship between surge height and flood damage on the U.S. coast is spatially variable and non-linear. Thus, to translate our projected surge heights into human impacts, we use CA-Surge, a simple bathtub-style inundation model with an overland frictional attenuation effect (as described in the Methods section). Note that since the CA-Surge model is only an approximation of true inundation physics, more emphasis should be placed on relative changes than on absolute totals.

\begin{figure}
    \centering

    \begin{subfigure}[]{0.79\textwidth}
        \centering
        \includegraphics[width=\textwidth]{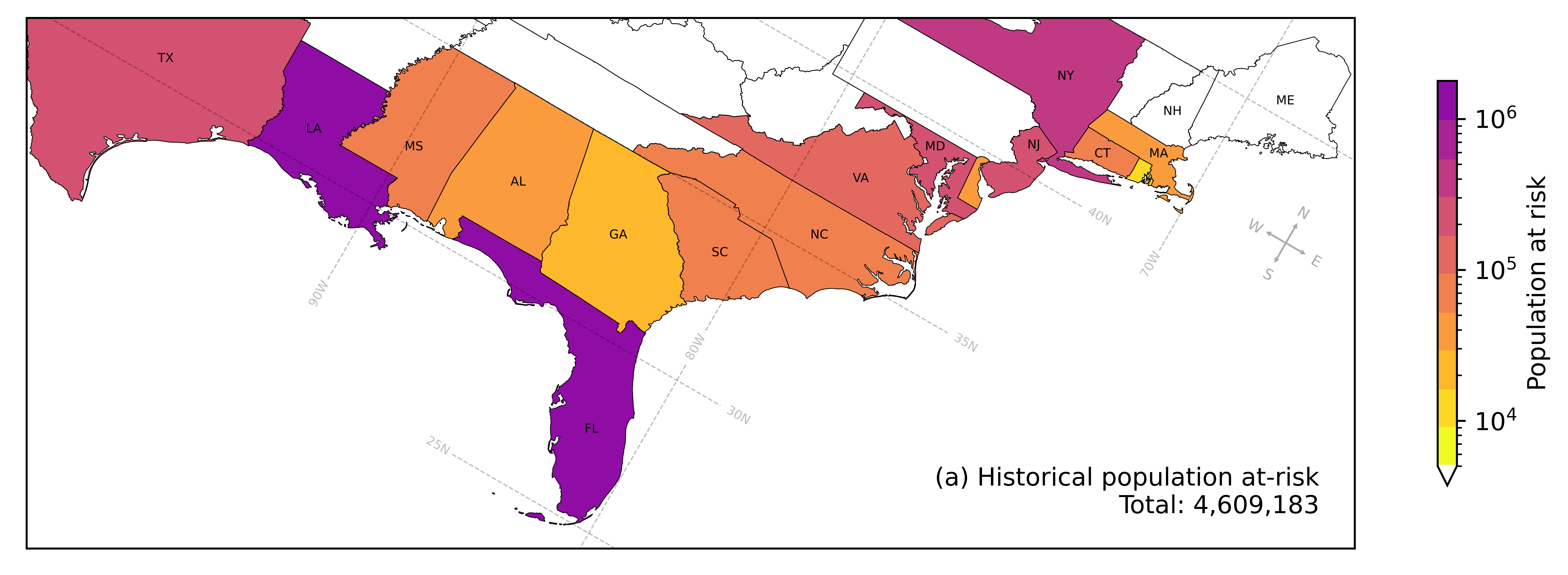}
        % \caption{Population at risk in the historical period}
        \phantomsubcaption
        \label{fig:inundation-hist}
    \end{subfigure}
    % \hfill
    \begin{subfigure}[]{0.79\textwidth}
        \centering
        \includegraphics[width=\textwidth]{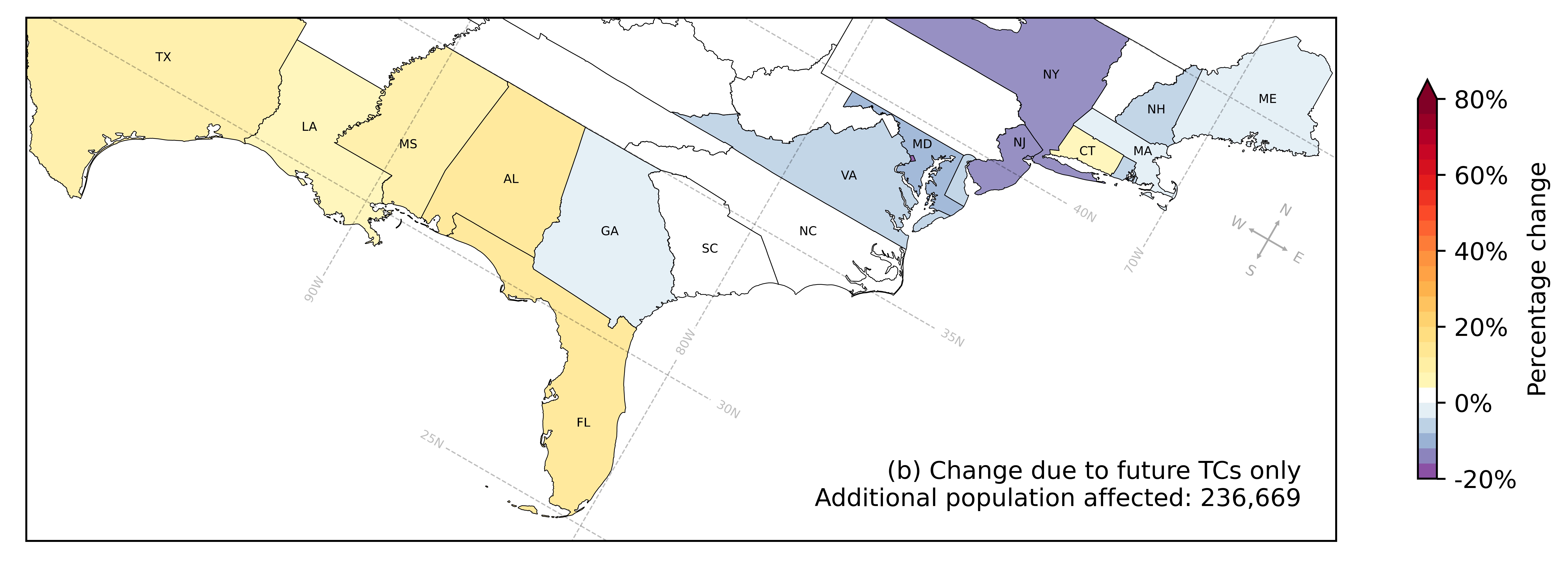}
        % \caption{Change in risk with future period TCs only}
        \phantomsubcaption
        \label{fig:inundation-futu-tcs}
    \end{subfigure}
    % \hfill
    \begin{subfigure}[]{0.79\textwidth}
        \centering
        \includegraphics[width=\textwidth]{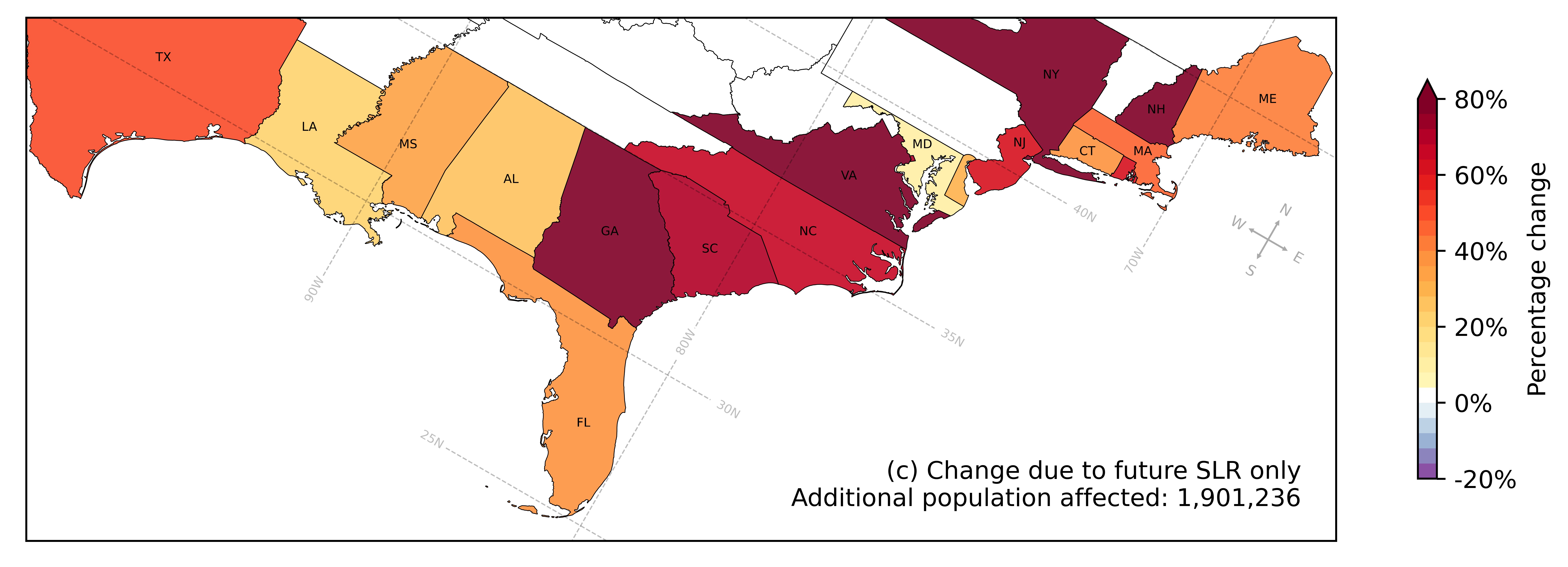}
        % \caption{Change in risk with future period sea-level rise only}
        \phantomsubcaption
        \label{fig:inundation-futu-slr}
    \end{subfigure}
    % \hfill
    \begin{subfigure}[]{0.79\textwidth}
        \centering
        \includegraphics[width=\textwidth]{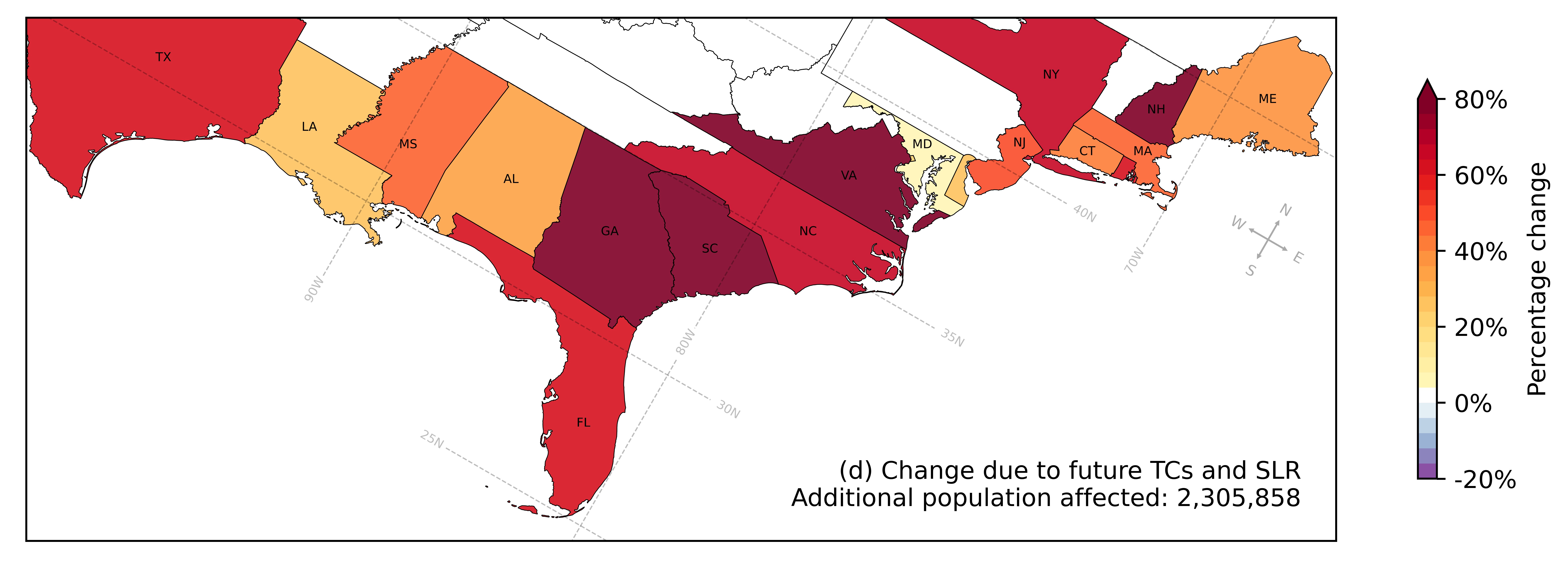}
        % \caption{Change in risk with future period TCs and sea-level rise}\
        \phantomsubcaption
        \label{fig:inundation-futu-tcs-slr}
    \end{subfigure}
    % \hfill

    \caption{Modeled population at risk from the 100-year flood event in (\textbf{a}) the historical period, and the percentage change in the future period with (\textbf{b}) future TCs only, (\textbf{c}) future sea-level rise (SLR) only, and (\textbf{d}) the combination of the two. All panels show the ensemble median estimate.}
    \label{fig:inundation}
\end{figure}

% inundation results
Even under historical climate conditions, the flooding risk posed by the 100-year storm surge event is significant---with an estimated 4.6 million people at risk---yet this risk is substantially heightened in our projected future scenario (Fig. \ref{fig:inundation}). Similar to the changes in surge heights, future changes in inundation risk are driven primarily by sea level rise ($+1.9$ million people at risk above historical; Fig. \ref{fig:inundation-futu-slr}) with changes in future TC behavior being a secondary contributor ($+0.24$ million people at risk above historical; Fig. \ref{fig:inundation-futu-tcs}). Taken together, risk is projected to increase in every coastal state (Fig. \ref{fig:inundation-futu-tcs-slr}), with a 50\% increase in population at risk overall ($+2.3$ million people above historical). Florida alone projects to see an additional roughly one million residents at risk in the future period, as it sees one of the larger state-level relative increases in risk on top of its already large historical risk.

% Inundation results

\begin{figure}
    \centering

    \includegraphics[width=0.9\textwidth]{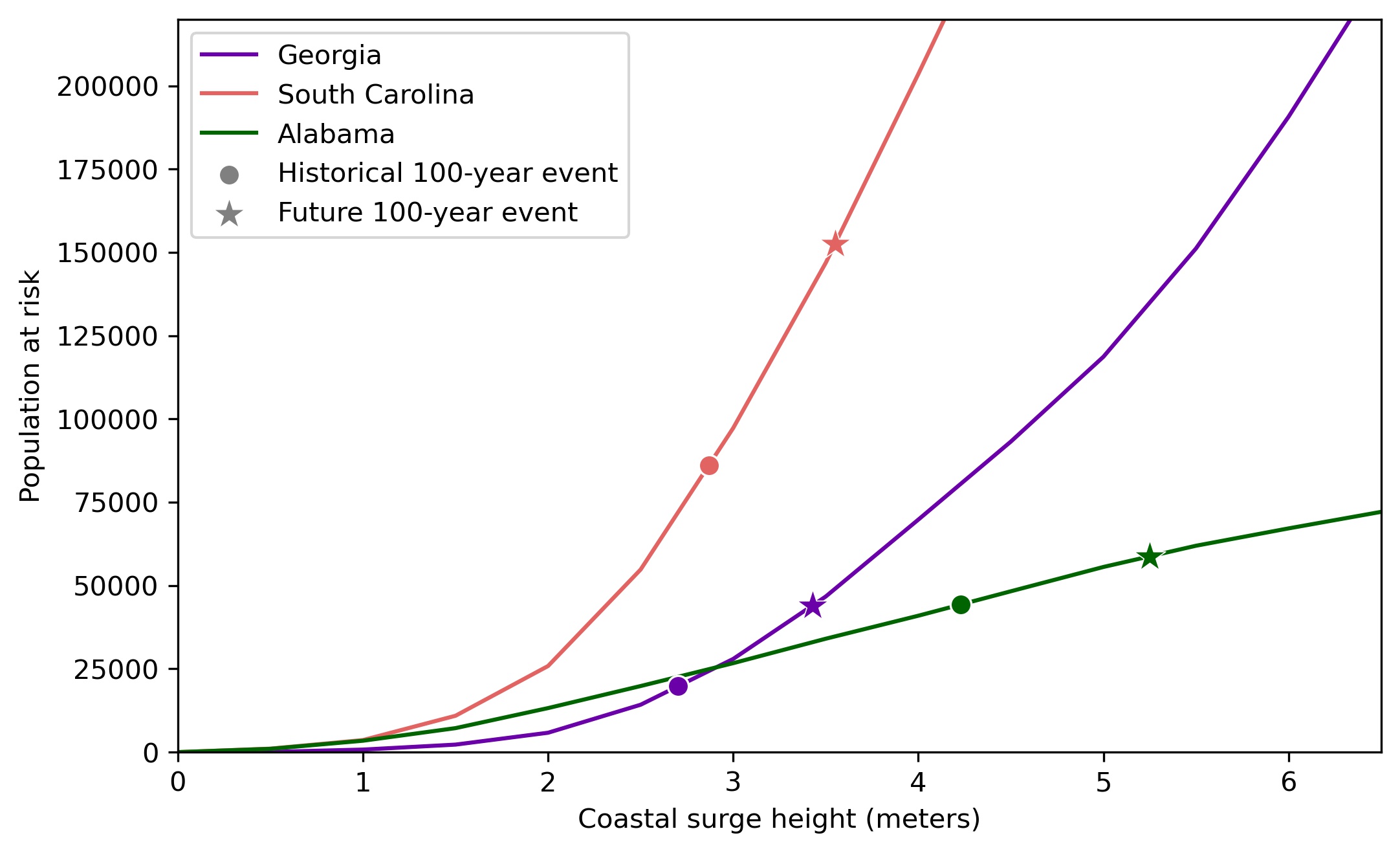}

    \caption{Curves relating average coastal surge height to population at risk for three selected states. The rapid steepening of Georgia and South Carolina's risk curves leads to large future changes in inundation risk, despite these states having both a lower historical 100-year surge height and a smaller absolute change in future surge height than states such as Alabama.}
    \label{fig:ga-sc-al-risk}
\end{figure}

% inflection points and uncertainty
Crucially, the percentage change in inundation risk at a state level is not always proportional to the local increase in 100-year surge height. While the largest future changes in surge height were observed in the central Gulf states of Louisiana, Alabama, and Mississippi, relative changes in coastal flood risk are most concentrated along the southeast Atlantic coast. For example, although Georgia and South Carolina see relatively moderate increases in 100-year surge heights in the future compared to many other states (Fig. \ref{fig:deepsurge-futu-change-slr}), they have among the largest percentage increases in inundation risk (Fig. \ref{fig:inundation-futu-tcs-slr}). This suggests that these states' coastal topographies may exhibit critical thresholds, above which population centers currently insulated from risk may become rapidly more vulnerable. This hypothesis is explored in Figure \ref{fig:ga-sc-al-risk}, which plots the relationship between average coastal surge height and population at risk for Georgia and South Carolina, with Alabama as a counterexample. Around the two meter threshold of coastal surge height, Georgia and South Carolina's risk curves inflect upward sharply, while Alabama's remains fairly linear. Thus, despite Alabama exhibiting a much larger future change in surge height on top of an already higher baseline 100-year surge, Georgia and South Carolina experience larger increases (in both absolute and percentage terms) in population at risk. These drastic differences in risk curve slope indicate that any future increase in surge height for Georgia and South Carolina is much more damaging than the same increase in Alabama would be. This finding underscores the importance of framing storm surge projections in terms of inundation risk instead of surge height, as these impactful nonlinear relationships would otherwise be missed.

% Inundation results

The large sample size of synthetic TC events and ensemble forcings allow for robust uncertainty quantification as well, which also exhibits nonlinear and heterogeneous patterns. The spatial distribution of uncertainty (Supplementary Fig. 3)---arising from the combination of uncertainties in GCM-induced TC behavior and sea-level rise---varies widely, with some states seeing proportionally much wider 90\% uncertainty bounds than others (e.g., New York vs. Louisiana), or greatly increasing uncertainty in the future (e.g., Florida). Beyond the 90\% bounds, the tails of the distributions tend to be much longer in the future than in the historical period, due to long tails in the sea-level rise projections (Supplementary Fig. 4). Very generally, uncertainty in population at risk from the 100-year event appears to be lower in the Gulf, and larger along the Atlantic coast.

\section{Discussion}
% summary
This study couples an ensemble of global projections of future ocean-atmospheric conditions, a synthetic TC model, a novel deep learning storm surge model, and an efficient inland inundation model to assess changes in storm surge risk and associated flooding impacts for the U.S. coastline. With an unprecedented sample size of TC events, we find substantial ($\sim 50$\%) increases in population at risk of once-in-a-century surge flooding, and characterize the spatial pattern and variability of this risk. While increases in 100-year surge heights are most prominent along the Gulf Coast, the pattern of inundation risk (as modulated by coastal topography and population densities) is revealed to be much more variable, exhibiting complex and non-linear interactions.

% future work
Our deep learning storm surge model, DeepSurge, is not a replacement for numerical storm surge models but serves as a useful complement to them, particularly in situations where traditional models may pose an intolerable computational cost or throughput. While this study demonstrates the utility of such a deep learning storm surge model in assessing surge risk, the technique is still evolving, and---like all data-driven methods---is inherently limited by the quality and scope of its training data. Given that the model operates on individual points rather than a fixed mesh, we intend to explore assimilating additional data from diverse regional and high-resolution hydrodynamic models during the training process. This may help address known biases in complex regions such as the Chesapeake Bay, where resolving fine-scale dynamics remains challenging. A similar technique may further allow the expansion of our model to new basins and domains across the globe.

% impacts and inequities of flooding
While accurately projecting flood risk is important on its own, continuing study is needed to further out understanding of the broader socioeconomic impacts of flooding. Beyond the immediate threats to wellbeing, floods cause a wide range of adverse and long-lasting effects, including increased disease and mortality \citep{alderman_floods_2012, stephens_excess_2007}, mental health disruptions \citep{graham_flood-_2019, alderman_floods_2012}, and unemployment \citep{allaire_socio-economic_2018, peek-asa_flood-related_2012}. Complex and varied patterns in the socioeconomic effects of flooding emerge across different coastal regions \citep{montgomery_social_2013, maldonado_exposure_2016, qiang_disparities_2019, herreros-cantis_shifting_2020, smiley_social_2022}, particularly in its impacts to housing and home valuations \citep{varela_varela_surge_2023, zhang_flood_2016, van_der_straten_flooded_2023, billings_let_2022}. Given these ramifications, careful consideration of adaption, mitigation, and retreat strategies at local and national scales is essential \citep{neumann_joint_2015}.

% \clearpage

% \acknowledgments

\section*{Acknowledgments}
% https://www.nersc.gov/users/accounts/user-accounts/acknowledge-nersc/
This work was supported by the Multisector Dynamics and Regional and Global Model Analysis program areas of the U.S. Department of Energy (DOE), Office of Science, Office of Biological and Environmental Research as part of the multi-program, collaborative Integrated Coastal Modeling (ICoM) project. The Pacific Northwest National Laboratory is operated for DOE by Battelle Memorial Institute under contract DE-AC05-76RL01830. 

This research used resources of the National Energy Research Scientific Computing Center (NERSC), a U.S. Department of Energy Office of Science User Facility located at Lawrence Berkeley National Laboratory, operated under Contract No. DE-AC02-05CH11231 using NERSC award BER-ERCAP0024320. This work also used the computing resources of Pacific Northwest National Lab’s Institutional Computing facility.

\section*{Data Availability}
DeepSurge-predicted storm surge heights for all 900,000 synthetic storm tracks are made publicly available on Zenodo: \url{https://doi.org/10.5281/zenodo.15021868}. All other data sources used in this study are publicly available, as is the ADCIRC algorithm.

\bibliographystyle{icml2023}
\bibliography{Surge}

\end{document}

\maketitle

\section{RAFT synthetic tropical cyclones}
The Risk Analysis Framework for Tropical Cyclones (RAFT) \citep{xu_deep_2021, balaguru_increased_2023, xu_north_2024} is a synthetic TC downscaling method. It uses a random seeding process for cyclogenesis conditioned on historical observations, with a physical beta-advection model for track propagation \citep{emanuel_statistical_2006, marks_beta_1992, kelly_shape_2018} and a deep learning regressor to model 6-hourly intensity changes over the lifetime of a storm \citep{xu_deep_2021}. Radius of maximum wind is parameterized with a log-transformed linear regression on latitude and maximum wind speed, following \citet{willoughby_parametric_2006}. Since large uncertainties remain about the rate of TC genesis in the future \citep{knutson_tropical_2020, murakami_patterns_2022, chavas_tropical_2024}, we use a fixed rate of 14.91 seeds per year (the observed historical rate in IBTrACS from 1980-2014) for both the historical and future periods. Note that the intensity model quickly dissipates or sustains TC seeds depending on how favorable their environment is, which allows the climate to implicitly regulate TC frequency. 

Climate conditions are gathered for the historical (1980--2014) and end-of-century future (2066--2100) periods under Shared Socioeconomic Pathway SSP5-8.5 from nine CMIP6 models: the Euro-Mediterranean Centre on Climate Change coupled climate model (CMCC-CM2-SR5), Canadian Earth System Model (CanESM5), Energy Exascale Earth System Model (E3SM), EC-Earth Consortium Model (EC-Earth3), Geophysical Fluid Dynamics Laboratory Climate Model (GFDLCM4), Institute Pierre Simon Laplace Climate Model (IPSL-CM6A-LR), Model for Interdisciplinary Research on Climate (MIROC6), Max Planck Institute Earth System Model (MPI-ESM1-2-LR), and Meteorological Research Institute Earth System Model (MRI-ESM2-0). We generate 50,000 TCs from the forcings of each of the 18 CMIP6 model-scenario pairs, for a total of 900,000 tracks. Due to climatological biases in the CMIP6 ensemble, we apply bias correction to the intensities of the synthetic TCs. Bias correction is implemented with a spatially-aware quantile delta mapping \citep{cannon_bias_2015} to align the distribution of historical TC intensities with historical observations, and preserve differences between corresponding quantiles in historical and future scenarios, as in \citet{lipari_amplified_2024} and \citet{rice_projected_2025}.

\section{Storm surge modeling}

\subsection{ADCIRC simulations}
The numerical model selected for generating the training dataset for DeepSurge is ADCIRC \citep{luettich_jr_solution_1991, luettich_adcirc_1992, pringle_global_2021} v53.04. ADCIRC is an ocean circulation model that has been widely used in storm surge studies. The computational grid used in this study was based on an unstructured mesh developed by \citet{dietrich_hurricane_2011}\footnote{Available at: \url{https://ccht.ccee.ncsu.edu/example-input-files-for-swanadcirc/}}. It spans the U.S. East Coast, the Gulf of Mexico, and the Caribbean Sea, with the open boundary defined along the North Atlantic Ocean. The horizontal resolution of the original mesh was relaxed to 15,467 nodes to balance computational costs and accuracy of the simulations, with a final resolution of approximately 25 km along coastlines and a 150 km resolution at the open boundary (Fig. \ref{fig:adcirc-grid-and-storms}a).

Historical tropical cyclone data for the North Atlantic was retrieved from NOAA’s International Best Track Archive for Climate Stewardship (IBTrACS) \citep{knapp_international_2010, knapp_international_2018}. The North Atlantic IBTrACS dataset was subsampled to include only tracks contained within the extents of the ADCIRC model domain and a minimum storm length of 3 days, allowing for data gaps in wind speed and atmospheric pressure records of no more than six hours (gaps shorter than six hours were filled through linear interpolation). 279 tracks met these criteria (Fig. \ref{fig:adcirc-grid-and-storms}b). Wind fields were generated following the methodology outlined by \citet{emanuel_self-stratification_2011} using each track’s maximum wind speed and radius of maximum wind, and a maximum radius of storm influence set at 300~km. Pressure fields were generated following the methodology presented by \citet{holland_revised_2008}. No other forcings (e.g., tidal) were included in the ADCIRC simulations. To ensure the stability of the model, storm events shorter than 5 days were simulated using a ramp function to extend the total simulation time to 5 days (e.g., a 2-day ramp period for a 3-day storm event). Each simulation was executed with a 1-second time step and the resulting water levels (i.e., storm surge) were stored at 1-hour intervals over the entire computational domain.

\subsection{DeepSurge architecture \& training}
This section assumes some basic familiarity with neural network architectures. For an introduction to neural networks we recommend \citet{lecun_deep_2015}, and \citet{emmert-streib_introductory_2020} for a more in-depth exploration.

The basic structure of the neural network, as detailed in Main Text Figure 1, has four stages: 1) encode the timeseries and spatial data separately, 2) convert them to compatible shapes and concatenate them together, 3) apply a Long Short-Term Memory (LSTM) \citep{hochreiter_long_1997} layer to understand the temporal development, and 4) decode the features to predict a maximum surge level. To compactly notate tensor shapes, let $T$ denote the (variable-length) time dimension, $I$ denote the number of input features in the timeseries, and $S$ denote the number of spatial images. The timeseries ($T$x$I$) is encoded with a pair of one-dimensional convolutional layers with a kernel size of one, to expand the series to shape ($T$x256). The spatial stack (64x64x$S$) is encoded with four successive 2-dimensional convolutional layers with striding and pooling to downsample to a vector (1x256) describing the spatial information. We then repeat this vector $T$ times to create a tensor ($T$x256) that can be concatenated with the temporal features to create a $T$x512 tensor. This combined feature is passed through an LSTM layer---a recurrent sub-network that processes each step of a timeseries in the context of information remembered from all previous steps---which outputs a new sequence ($T$x256). Lastly, we decode these LSTM outputs with another series of 1-dimensional convolutional layers, resulting in a final output sequence ($T$x1). Taking the maximum over the $T$ dimension returns the scalar maximum surge prediction for the storm at that given node. The network is constructed and trained in TensorFlow/Keras \citep{martin_abadi_tensorflow_2015}, and has a total of 1.7 million parameters.

Prior to training, 25\% of storms in the dataset are set aside for testing, and never trained on. An additional 15\% of remaining storms are used as the validation set, to test for overfitting during the training process. Since extreme surge levels are of more interest and simultaneously more rare in the training data, two adjustments are made: Each example's loss is weighted proportional to the ADCIRC peak surge height in meters plus one (e.g., a surge of 6~m has a weight of 7, while a surge of 0~m has a weight of 1), and examples with ADCIRC peak surges below 1 meter are sampled less frequently to correct for their over-representation in the dataset. Training occurs in epochs, each consisting of 500 batches with a batch size of 32. During training, low levels of zero-centered Gaussian noise ($\sigma=$1e-3) are added to the inputs as a simple regularization mechanism. We use the Adam optimizer \citep{kingma_adam_2014}, and the mean-squared error between the maximum neural network prediction and maximum ADCIRC modeled surge as the loss target. At the end of each epoch, the loss on the validation set is assessed, and the best model iteration according to this metric is saved. The learning rate begins at 2e-3, and is multiplied by a factor of 0.2 after any three consecutive epochs with no validation loss improvement. Training is stopped after six consecutive epochs with no validation loss improvement. This architecture was developed and tuned over numerous manual iterations, including experimenting with learning rates, batch sizes, alternate model structures, batch normalization, dropout, scaling the width and depth of the network, different recurrent units, and regularization techniques.

\subsection{DeepSurge computational efficiency}
For an approximate comparison of the computational speedup provided by our method, we predict Hurricane Katrina's storm surge using both DeepSurge and ADCIRC on high-performance computing systems. We find DeepSurge to be approximately 12.5x faster in terms of CPU-hours, with an upper bound of 96x when predicting hundreds of storms in sequence (since DeepSurge's initialization stage is the bulk of the computational burden, which can subsequently be shared across multiple simulations). Even with this significant speedup, the computational expense of simulating 900,000 storms is not insignificant; approximately 350 high-performance compute node-hours were utilized (with 256 cores per node, $\sim90,000$ core-hours).
% ~350 node hours, 14 wall clock days

\subsection{DeepSurge validation \& sensitivity analysis}
DeepSurge demonstrates promising generalization skill on the test set, which consists of 71 storms that the model has never been trained or tuned on. When compared to the corresponding ADCIRC simulations, DeepSurge achieves an 81.5\% $R^2$ score, a mean squared error of 0.224~m, and a mean absolute error of 0.258~m in predicting peak surge heights per node. As error metrics in terms of raw surge height are difficult to contextualize on their own, we later assess the sensitivity of our inundation results to the model's biases and errors.

To compare with observed storm surge data, we perform a validation against NOAA tide gauge observations\footnote{Accessible from the Tides \& Currents API: \url{https://api.tidesandcurrents.noaa.gov/mdapi/prod/\#}}. All available peak tide gauge measurements within a 300~km radius of each test-set storm's track are collected and co-located with the predictions from the nearest ADCIRC/DeepSurge node. Peak surge for each tide gauge is calculated as the maximum difference between observed water level and expected tide. Only verified data with gaps of less than one hour during the lifetime of the storm are considered, resulting 194 valid gauge observations. Although this is highly accurate data, it is notably incomplete; tide gauges often fail or malfunction during extreme surge events, with failure observed at surge heights as low as 1.22~m (see \citet{needham_data-driven_2014}'s Figure 5.1). This effect causes a fairly strong sampling bias toward smaller surge levels, with the largest observed surge in this collection being 3.1~m despite the fact that a number of storms in the sample (including 2005's Hurricane Katrina) are known to have generated much larger surges. Gauge observations may be influenced by factors not modeled by ADCIRC or the neural network, such as rainfall, river discharge, and background non-cyclonic winds, which makes them an imperfect comparison for our surge-only ADCIRC and DeepSurge simulations, though we expect these influences to be small relative to TC-driven surge in most cases. DeepSurge shows reasonable skill in capturing these gauge-observed surge peaks, with error metrics ($R^2=0.403$, $\text{MAE}=0.474$m, $\text{RMSE}=0.664$m) similar to the ADCIRC simulations ($R^2=0.427$, $\text{MAE}=0.543$m, $\text{RMSE}=0.882$m). Pearson and Spearman correlations are highly significant ($p \ll 0.001$). Possibly due to the negative sampling bias of the gauge observations, both DeepSurge and ADCIRC exhibit positive mean biases ($+0.217$m and $+0.203$m respectively).

Additionally, a sensitivity analysis is performed to understand how the biases of the DeepSurge model may impact our final inundation estimates. Specifically, we estimate the bias at each DeepSurge node as the mean of the biases at all gauge observations within three degrees of the node, weighted inversely by distance. This process is averaged over 200 bootstrap samples of the gauge bias observations for robustness. The derived bias (Fig. \ref{fig:deepsurge-bias}) is generally positive as noted prior, with near zero bias ($<0.25$~m) in the western Gulf of Mexico and eastern Florida, moderate bias (generally $<0.5$~m) in the rest of the Gulf and Southeast, and larger biases in the Northeast---though the sample size in that region is notably less robust, with 20 total gauge-prediction pairs north of $37\degree N$ caused by only 4 storms. Correcting for this estimated bias in DeepSurge's 100-year surge event and recalculating inundation risk results in an 18\% reduction in historical-period population affected, and a 14\% reduction in future-period population affected. Use of these bias-corrected totals actually indicates an even stronger relative change in inundation risk in the future of $+57\%$, compared with the $+50\%$ found using the uncorrected totals. At the individual state level, all states in the Gulf and Southeast see corrections of under $\pm30\%$ (and all but Georgia and South Carolina under $\pm20\%$), with the near-zero biases in Texas. Corrections are much larger in the Northeast, exceeding $-50\%$ in Massachusetts and New Jersey, though again we note the much lower confidence in estimated biases there. Given the limited sample size, incompleteness of the tide gauge records, and confounding factors previously discussed, we treat these results as only rough indications of the directions and magnitudes of DeepSurge biases. For these reasons, and because the uncorrected total actually present a more conservative view of relative future changes in risk, we choose to report the uncorrected totals in the main manuscript.

\subsection{Validation of modeled 100-year surge height}

To quantitatively verify that the combination of DeepSurge and RAFT are producing accurate 100-year surge heights, we compare our ensemble median return levels with the historical 100-year event observational estimates from \citet{needham_data-driven_2014}, who undertook an exhaustive analysis of the historical storm surge record 1900--2014 throughout the Gulf of Mexico, and three of independent surge modeling techniques: \citet{gori_tropical_2022} forced a similar synthetic TC model with NCEP reanalysis for a historical period (1980--2005) and eight CMIP6 GCMs for a future period (2070--2100 under SSP5-8.5) to model the 100-year storm \textit{tide} above mean higher high water (MHHW) with a peaks-over-threshold approach using ADCIRC. \citet{muis_global_2023} forced the Global Tide and Surge Model (a hydrodynamic model based on Delft3D) with historical ERA5 reanlysis (1985--2014) as well as five high-resolution CMIP6 models for a historical (1950--2014) and near future (2015--2050 under SSP5-8.5) period, using a peaks-over-threshold approach to estimate 100-year storm surge levels. Lastly, for a comparison with a simple parametric model we generate surges with the Storm Surge Hazard Potential Index (SSHPI) developed by \citet{islam_new_2021}, forced by the same RAFT synthetic TCs as DeepSurge. The historical 100-year estimates from each of the modeling methods are presented in Fig. \ref{fig:modelcomp-100yr-events}.

The 100-year estimates from \citet{needham_data-driven_2014}\footnote{Data available in \citet{needham_data-driven_2014}'s Table 5.9}, \citet{gori_tropical_2022}\footnote{Data available at: \url{https://doi.org/10.17603/ds2-gv07-kf03}}, and \citet{muis_global_2023}\footnote{Data available at: \url{https://doi.org/10.24381/cds.6edf04e0}} are all directly published with no additional computation performed on our part, while Storm Surge Hazard Potential Index (SSHPI) \citep{islam_new_2021} estimates were derived using the following method: Predictions were computed by applying the SSHPI equations (see \citet{islam_new_2021}, their equations 4 and 5) to the same 900,000 synthetic TCs used to force DeepSurge. SSHPI-derived surge is computed at hourly timesteps for all nodes within 0.5 degrees of the storm center, and the maximum predicted surge at each node is saved. Because the SSHPI formulation requires a measure of the mean radius of 50-knot winds ($R_{50kt}$, in nautical miles) which is not directly modeled in our synthetic TCs, a linear regression on maximum wind speed ($V_{max}$, in knots), latitude (degrees North), and radius of maximum wind ($R_{mw}$, in nautical miles) was derived from all available observations ($n=222$) in the HURDAT database \citep{jarvinen_tropical_1984}. This simple formulation achieves a Pearson's correlation of $0.83$ and $R^2$ of $0.68$:
\begin{equation*}
    R_{50kt} = 0.596 \cdot V_{max} + 0.853 \cdot R_{mw} + 2.074 \cdot \text{lat} -69.044
\end{equation*}
The three methods' predictions are all co-located to the nearest node within 0.15 degrees, which enables consistent comparison between methods at 651 locations along the U.S. coast.

% compare magnitude against sparse Needham estimates 
The analysis by \citet{needham_data-driven_2014} reveals 100-year surge levels ranging from 2.53~m (Cedar Key, Florida) to 7.95~m (Bay St. Louis, Mississipi). Naturally these observations are subject to a great deal of random variability, as the chance occurrence or non-occurrence of particular extreme events (e.g., Hurricane Katrina) within this particular 115-year sample will bias the estimates; still, they remain the best available approximation of the spread of 100-year surge levels in this region. Co-locating the \citet{needham_data-driven_2014} estimates with the other methods ($n=18$) reveals that DeepSurge achieves the most comparable cumulative distribution of 100-year surge magnitudes (Fig. \ref{fig:needham-cdfs}). However, all methods including DeepSurge have insignificant ($p>0.05$) Spearman correlations against the spatial pattern of the sparse Needham estimates.

% compare spatial pattern against other models
In addition to comparing against the historical records, each of these methods' 100-year return level estimates are cross-compared. Given the wide variability in magnitudes, Spearman's rank correlation is used as a scale-invariant measure of pattern similarity. As shown in Figure \ref{fig:modelcomp-100yr-events}, DeepSurge has the best Spearman correlation with every other method, for both the historical and future periods (excepting the high correlation between the \citet{muis_global_2023} variants). The larger sample of synthetic TCs utilized in our method may allow it to capture a more robust signal and be less subject to random noise, which enables stronger agreement with the three alternatives than they exhibit between themselves.

\section{Inundation modeling}

\subsection{CA-Surge algorithm}

CA-Surge is a simple ``bathtub-style'' inundation model, with frictional attenuation. Conceptually, the algorithm starts with a water surface elevation (surge height) in the open ocean, and fills inland pixel-by-pixel until the water is not able to flow any further. The water depth for a pixel is computed as the difference between the water surface elevation and the ground elevation, with the water surface elevation being lowered (attenuated) as it moves across a pixel, in accordance with its land cover class.

The algorithm uses 3 input rasters and an attenuation multiplier parameter. All three rasters must be the same size and aligned with each other. DEM-Raster contains a digital elevation model of the study domain. Each cell (pixel) gives the height above (or below) mean sea level (in meters). Surge-Raster contains a surge height (in meters) above mean sea level for all cells (pixels) where surge originates (usually over the open ocean near the coast). All other cells (pixels) contain 0. Attenuation-Raster contains the surge attenuation values. Each cell (pixel) contains the drop in surge (in meters) as water horizontally traverses the cell. For example, if the cell contains the value 0.01, it means that the surge will be reduced by 1 centimeter if the surge traverses the cell. Attenuation-Multiplier is a numerical value between 0 and 1 that is applied to the values in the Attenuation-Raster, in accordance with \citep{vafeidis_water-level_2019}. If the multiplier is 0, then no attenuation will be applied to the surge as it traverses the raster cells (pixels). If the multiplier is 1, then the full attenuation value will be subtracted from the surge as it traverses the raster cells. CA-Surge is then defined by the following algorithm:

\begin{algorithm*}
\caption{CA-Surge}

Four internal matrices that match the size of the input rasters are initialized:
\begin{itemize}
    \item $el$: contains the DEM-Raster data.
    \item $att$: contains the Attenuation-Raster data.
    \item $wse$: contains the water surface elevation. The cells are initialized to the values of the Surge-Raster.
    \item $h$: contains the height of the surge water above the DEM elevations. Each cell is initialized to $wse$ minus $el$. The cell is set to 0 if any of the following conditions is true for the cell: the $wse$ cell contains no data, the $el$ cell contains no data, or $wse$ minus $el$ is less than or equal to 0. Cells that are set to a positive, non-zero value are added to a list of cells to a check list.
\end{itemize}

The model then iterates through the following steps:
\begin{enumerate}
    \item If the check list is empty, processing is complete, and the $h$ matrix is output as a raster.
    \item For each cell in the check list (the ``current cell"), apply the following to each of the current cell's eight adjacent neighbors (the ``neighbor cell"):
    \begin{enumerate}
        \item Compute $waterSurface = wse – (att * distance * \text{Attenuation-Multiplier})$, where $wse$ and $att$ are from the current cell, and distance is 1.0 for vertical and horizontal neighbors and $sqrt(2)$ for diagonal neighbors.
        \item Compute $depth = waterSurface - el$, where $el$ is from the neighbor cell.
        \item If $depth$ is greater than 0 and $el$ is not empty and the neighbor cell has not already been processed, or $depth$ is greater than the $h$ of the neighbor cell:
        \begin{itemize}
            \item $wse$ of the neighbor cell is set to $waterSurface$
            \item $h$ of the neighbor cell is set to $depth$.
            \item The neighbor cell is added to the check list of the next iteration.
        \end{itemize}
    \end{enumerate}
\end{enumerate}

\end{algorithm*}

\subsection{Validation of inundation modeling}

Accurate observations of inundated area from historical storm surge events are generally difficult to come by. Some of the best available data to our knowledge are FEMA high-water marks (HMWs) surveyed following 2005's Hurricane Katrina by contractor URS Group \citep{urs_group_final_2006, urs_group_high_2006-1, urs_group_high_2006}. This data contains a larger sample of surveyed points than most other HWM surveys, provides HWM elevation, and crucially labels each mark as caused to wave action, riverine flooding, or sustained storm surge, which is not common in other surveys. This enables an accurate evaluation of the surge-only DeepSurge method without complications from wave run-up and riverine flooding. Although the most famous flooding from this event was in Louisiana, much of it was due to or compounded by failures in the levee system; since these factors are outside of the modeling abilities of DeepSurge and CA-Surge, we focus on Mississippi and Alabama in this comparison, which received substantial flooding as well. We estimate inundation from this data by linearly interpolating a 3d water level surface from all HWMs, subtracting USGS ground elevation \citep{danielson_global_2011} to determine height-above-ground, and filtering all pixels which are not hydrologically connected to the ocean (Fig. \ref{fig:fema-hwm-surge}). Ground elevation is computed as the midpoint between the median and minimum elevation in each pixel to account for water's tendency to take the lowest available path. The extent of flooding seems to be roughly bounded by the outer extent of high-water mark locations in Fig. \ref{fig:fema-hwm-surge} which suggests this method is capturing the underlying hydrodynamics that caused the marks. LandScan data from all inundation pixels is used to estimate inundated populations, which are then aggregated to census-tracts (Fig. \ref{fig:fema-hwm-inundation}). DeepSurge and CA-Surge applied to the best track of Katrina (Fig. \ref{fig:deepsurge-hwm-inundation}) reasonably captures the pattern of the HWM-derived results in terms of percentage of population affected at the tract level (Pearson $r=0.77$, Spearman $r=0.78$, $p$-values $\ll 0.001,$ $R^2=0.59$, $\text{MAE}=9.1$ percentage points).

To validate the combination DeepSurge and CA-Surge for the entire coastline, we compare our state-level estimates of historical population at risk from 100-year storm surge inundation with those from \citet{crowell_estimate_2010} (see their Table 1), who estimate the 100-year coastal flood risk from the extremely high-resolution Flood Insurance Rate Maps (FIRMs) produced by FEMA. These FIRMs are developed with detailed and manual process combining modeling, observations, expert survey, and community input. Because of the effort required, FIRMs in some areas may be decades old. Additionally, since the maps do not differentiate between causes of flooding, \citet{crowell_estimate_2010} manually estimate the separation of surge- and riverine-driven flooding to isolate the former. Lastly, these coastal 100-year flood zones are spatially joined with census block-group population estimates from the 2000 census. Notably, they assume that population density is distributed uniformly within block-groups. While these approximations are possible sources of error, this FIRM-based assessment still represents one of the best available estimates of 100-year surge risk for the whole US coastline. We find that our historical-period 100-year inundation estimate correlates very highly with Crowell et al.'s totals, with Pearson and Spearman correlations of 0.95 and 0.87 respectively ($p \ll 0.01$). Interestingly, our DeepSurge \& CA-Surge method shows a negative bias in population affected for most states (Fig. \ref{fig:crowell-comp})---most significantly for barely-affected states e.g. Pennsylvania---which runs counter to the positive biases found against tide gauge observations and high-water mark analysis in previous sections. These differences are perhaps due to the inclusion of wave action effects in the FEMA modeling, or inaccuracies in either (or both) of DeepSurge or Crowell's methodologies.

\section{Drivers of change}

% SSHPI-assisted 
It is difficult to assess the contributions of individual storm characteristics to DeepSurge-predicted surge heights due to the black-box nature of neural networks and the complex, nonlinear, and time-dependent nature of the underlying physics. However, simple parametric models such as SSHPI \citep{islam_new_2021, islam_tropical_2022} enable the evaluation of the approximate direction and magnitude of effect that each storm characteristic has on surge height. SSHPI predicts peak surge based on the multiplicative combination of four factors: storm intensity (maximum wind speed), storm size (radius of 50-kt winds), storm translation speed, and bathymetry (distance to 30 meter isobath). As Figure \ref{fig:sshpi-factors}a-b shows, SSHPI projects a much more homogeneous increase in surge risk across the US coastline than DeepSurge, though the two do broadly agree on the sign of the change south of $35^{\degree}$N, while agreement is more mixed in the Northeast. Disagreements may be largely explained by SSHPI's indifference to the direction of storm motion; in regions where storms tend to make landfall at indirect angles relative to the coastline (e.g. much of the Northeast, Fig. \ref{fig:hist-storm-direction}) or are projected to increasingly do so in the future climate (e.g. western Florida, \ref{fig:change-storm-direction}), SSHPI's formulation is less reliable. Nevertheless, since it is forced with the same set of TCs as DeepSurge, disassembling SSHPI into its components is instructive: Increasing storm intensities are a strongly positive contribution to surge height across the domain (Fig. \ref{fig:sshpi-vterm}), while slightly larger radii of 50-kt winds---due largely to increasing intensities---are further weakly positive (Fig. \ref{fig:sshpi-rterm}). Decreasing storm translation speeds in the future period result in lower surges on open coastlines, and slightly larger surges in bays and estuaries (Fig. \ref{fig:sshpi-sterm}). In total, increasing storm intensity dominates the trend across nearly the entire coastline (Fig. \ref{fig:sshpi-dominant}).

%
%
%%%%%%%%%%%%%%%%%%%%%%%%%%%%%%%%
%
% FIGURES 
%
%%%%%%%%%%%%%%%%%%%%%%%%%%%%%%%%
%
%
\clearpage

% tc intensity
\begin{figure}
    \centering

    \begin{subfigure}[]{0.49\textwidth}
        \centering
        \includegraphics[width=\textwidth]{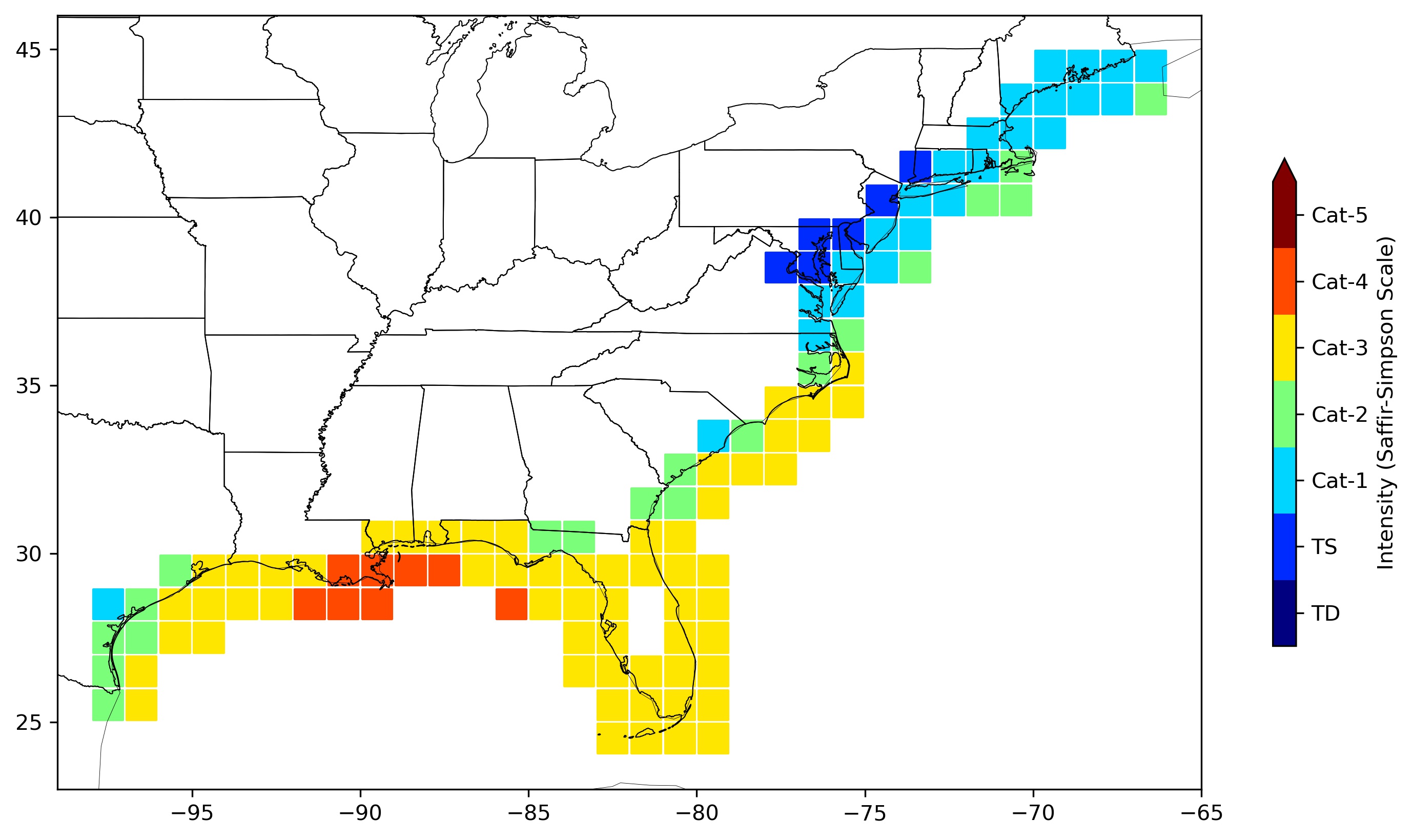}
        \caption{Historical 100-year TC intensity}
        \label{fig:hist-storm-intensity}
    \end{subfigure}
    \hfill
    \begin{subfigure}[]{0.49\textwidth}
        \centering
        \includegraphics[width=\textwidth]{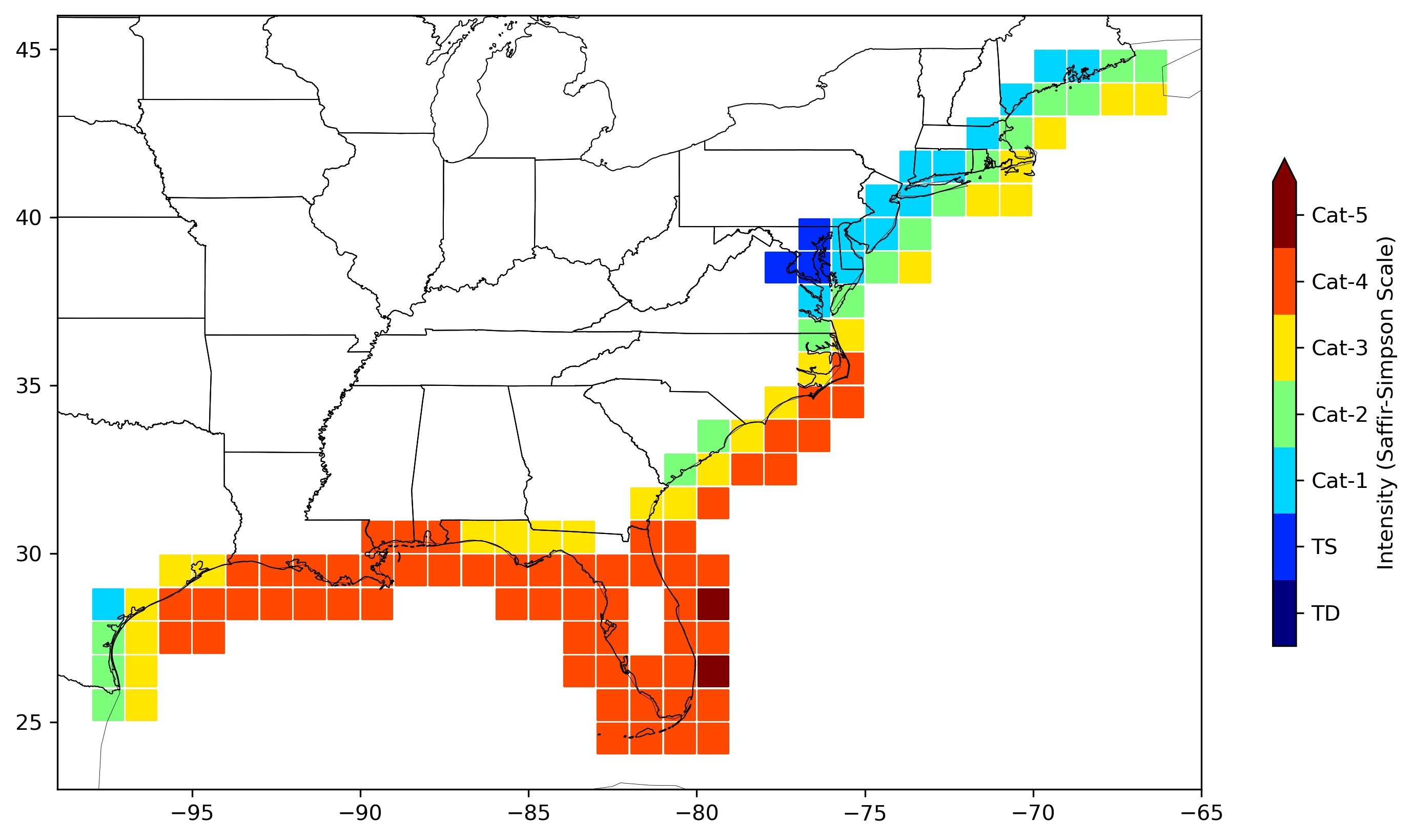}
        \caption{Future 100-year TC intensity}
        \label{fig:futu-storm-intensity}
    \end{subfigure}

    \begin{subfigure}[]{0.49\textwidth}
        \centering
        \includegraphics[width=\textwidth]{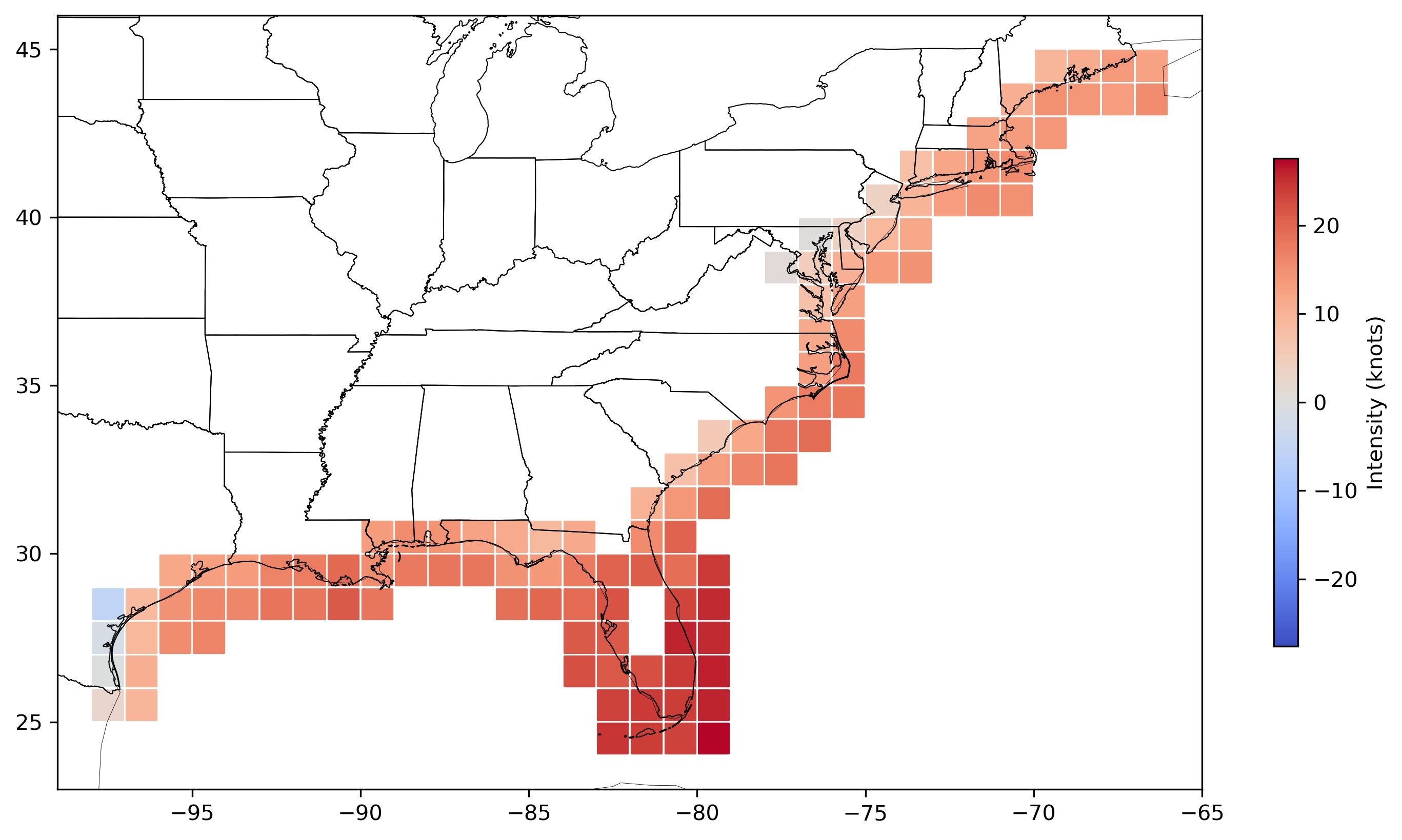}
        \caption{Future change in 100-year TC intensity}
        \label{fig:change-storm-intensity}
    \end{subfigure}

    \caption{Ensemble mean 100-year synthetic TC intensity for the (\textbf{a}) historical and (\textbf{b}) future periods, and (\textbf{c}) their difference.}
    \label{fig:storm-intensity}
\end{figure}

% average storm direction
\begin{figure}
    \centering

    \begin{subfigure}[]{0.49\textwidth}
        \centering
        \includegraphics[width=\textwidth]{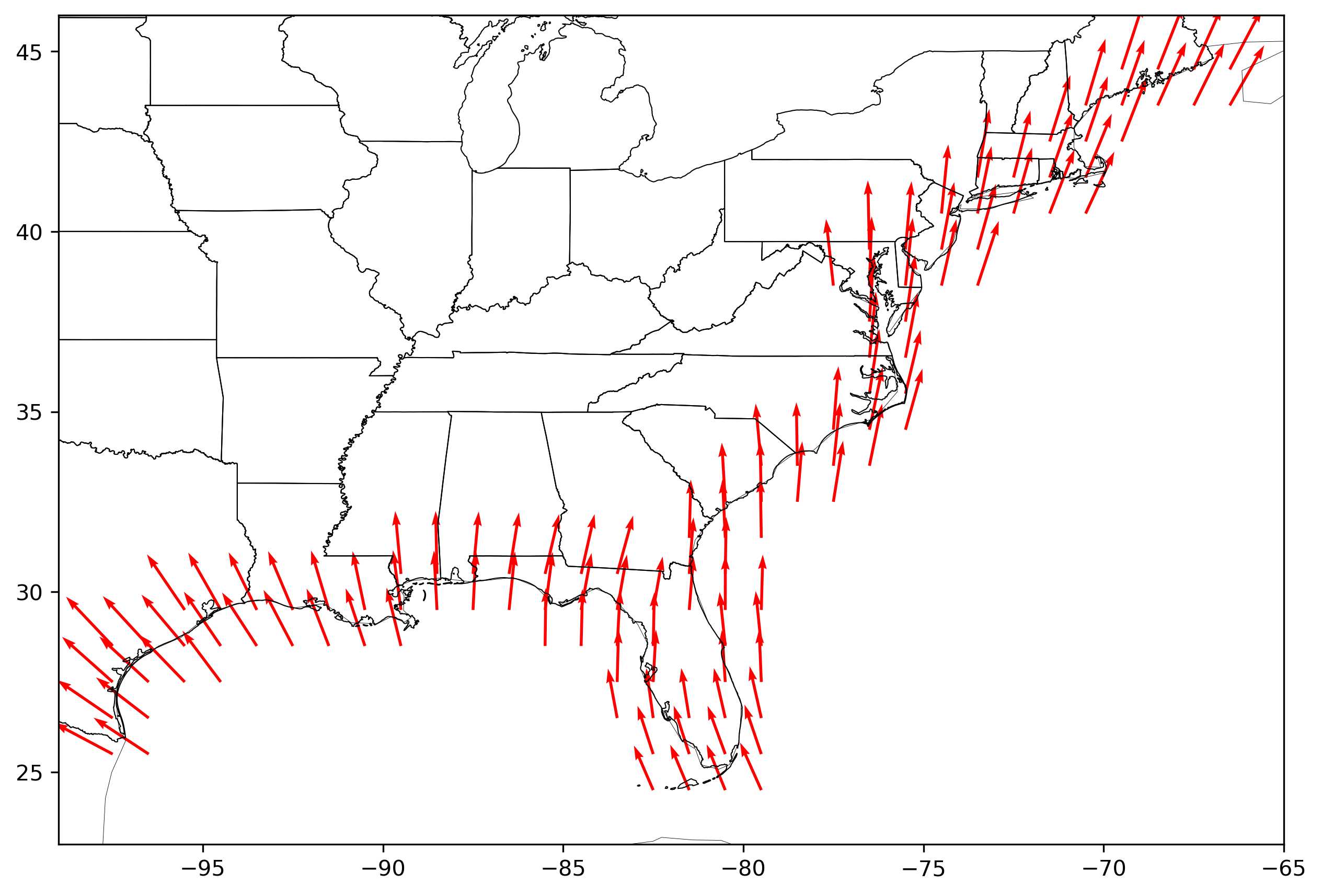}
        \caption{Historical storm direction}
        \label{fig:hist-storm-direction}
    \end{subfigure}
    \hfill
    \begin{subfigure}[]{0.49\textwidth}
        \centering
        \includegraphics[width=\textwidth]{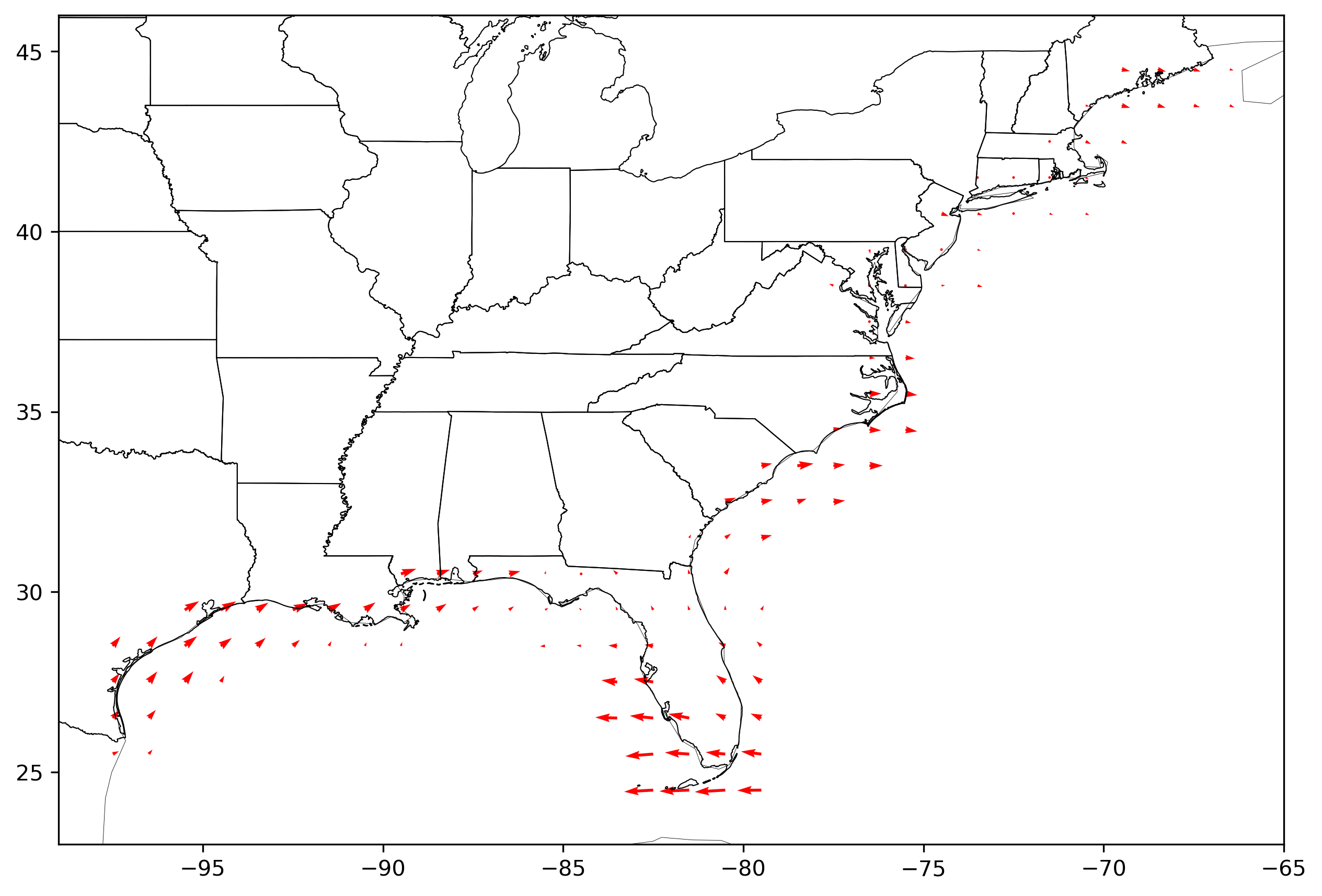}
        \caption{Future change in storm direction}
        \label{fig:change-storm-direction}
    \end{subfigure}

    \begin{subfigure}[]{0.49\textwidth}
        \centering
        \includegraphics[width=\textwidth]{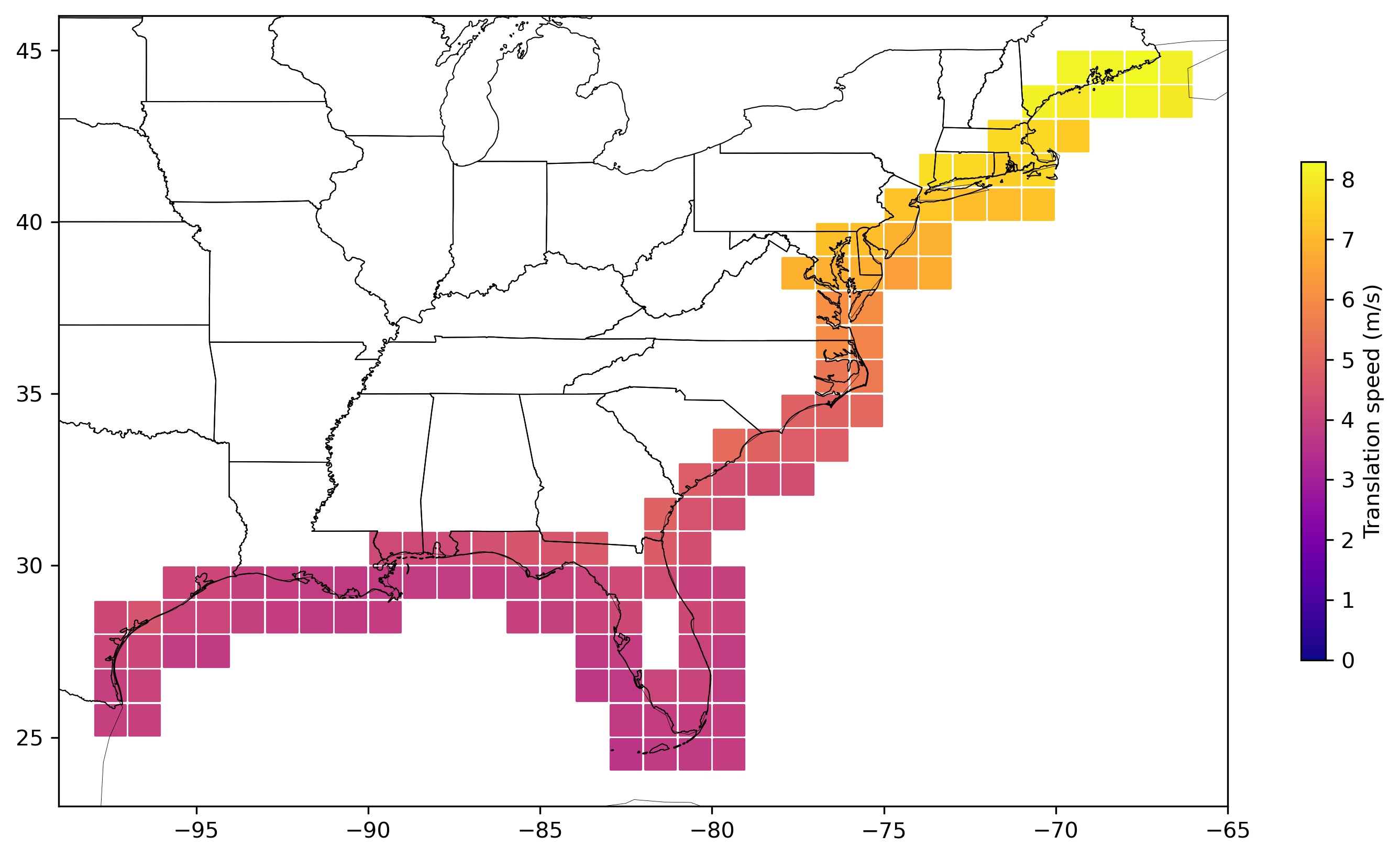}
        \caption{Historical storm translation speed}
        \label{fig:hist-storm-transspeed}
    \end{subfigure}
    \begin{subfigure}[]{0.49\textwidth}
        \centering
        \includegraphics[width=\textwidth]{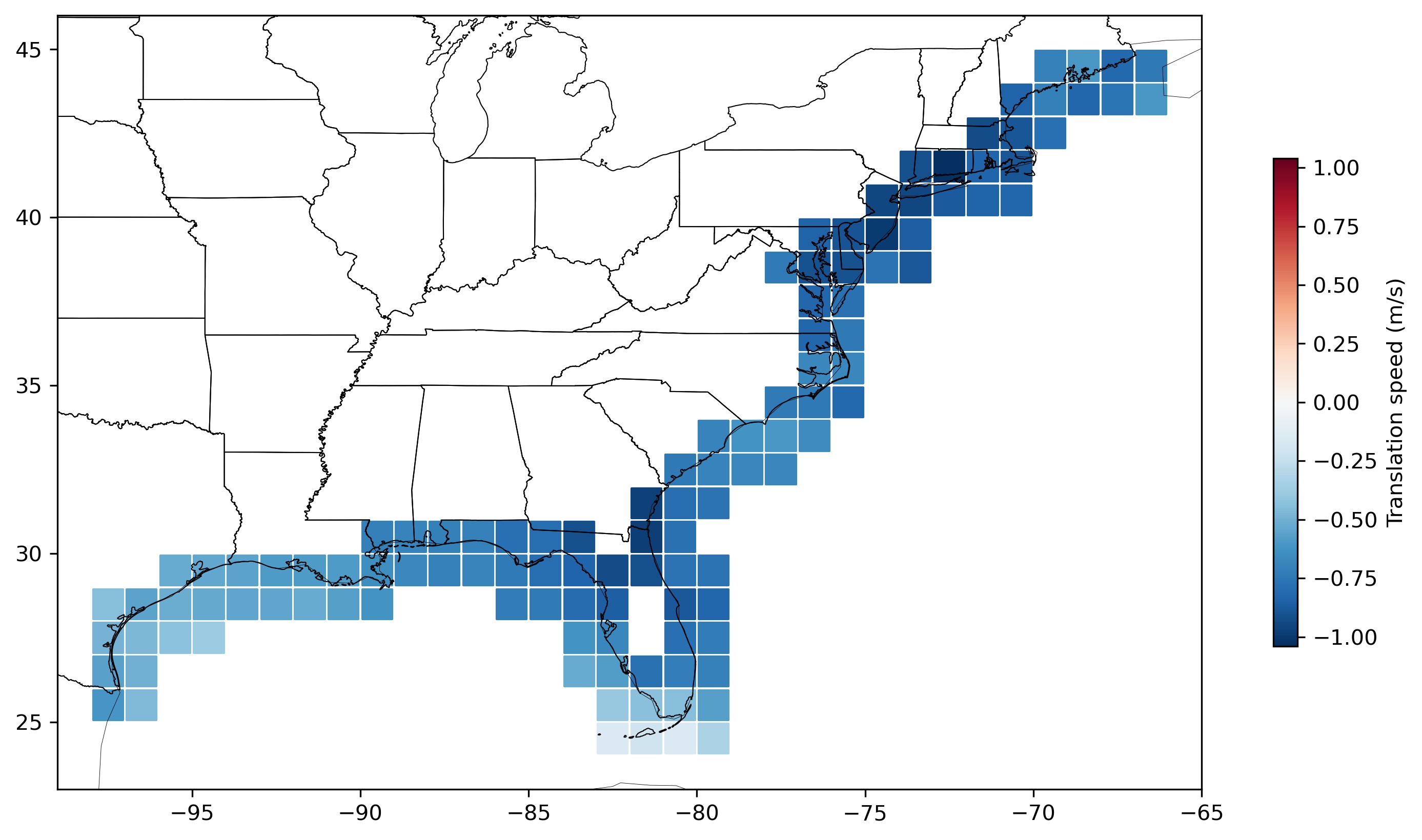}
        \caption{Future change in storm translation speed}
        \label{fig:change-storm-transspeed}
    \end{subfigure}

    \caption{(\textbf{a-b}) Average movement direction for all synthetic TCs (of at least Category 1 Saffir-Simpson strength) in the historical period weighted by magnitude of storm translation speed, and the future change. Arrows in panel \textbf{b} are 2x scaled relative to \textbf{a} for clarity. (\textbf{c-d}) Average synthetic TC movement speed in the historical period weighted by storm intensity, and the future change.}
    \label{fig:storm-direction}
\end{figure}

% Inundation uncertainty
\begin{figure}
    \centering

    \begin{subfigure}[]{0.49\textwidth}
        \centering
        \includegraphics[width=\textwidth]{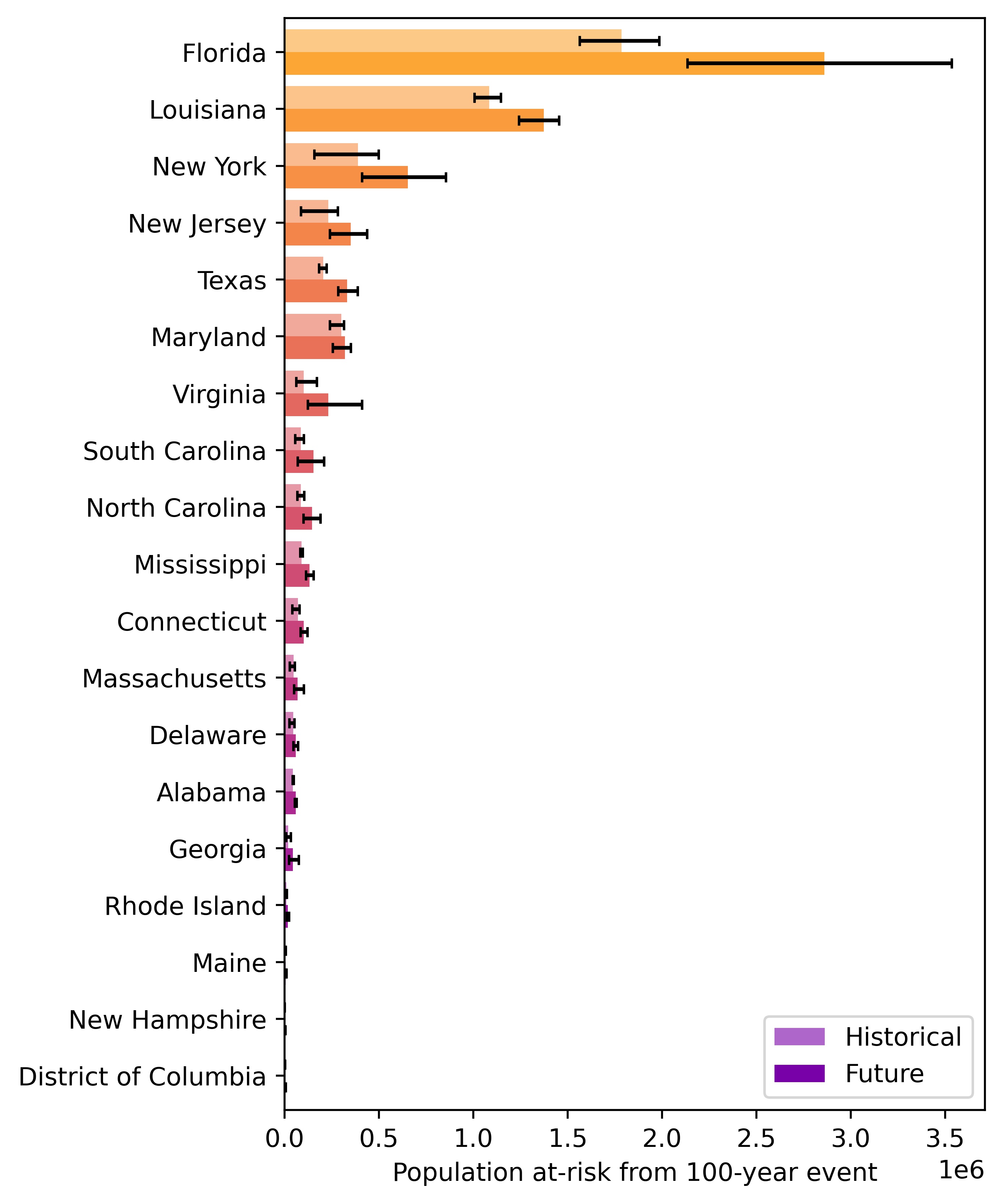}
        \caption{Linear scaling}
        \label{fig:inundation-uncertainty-unscaled}
    \end{subfigure}
    % \hfill
    \begin{subfigure}[]{0.49\textwidth}
        \centering
        \includegraphics[width=\textwidth]{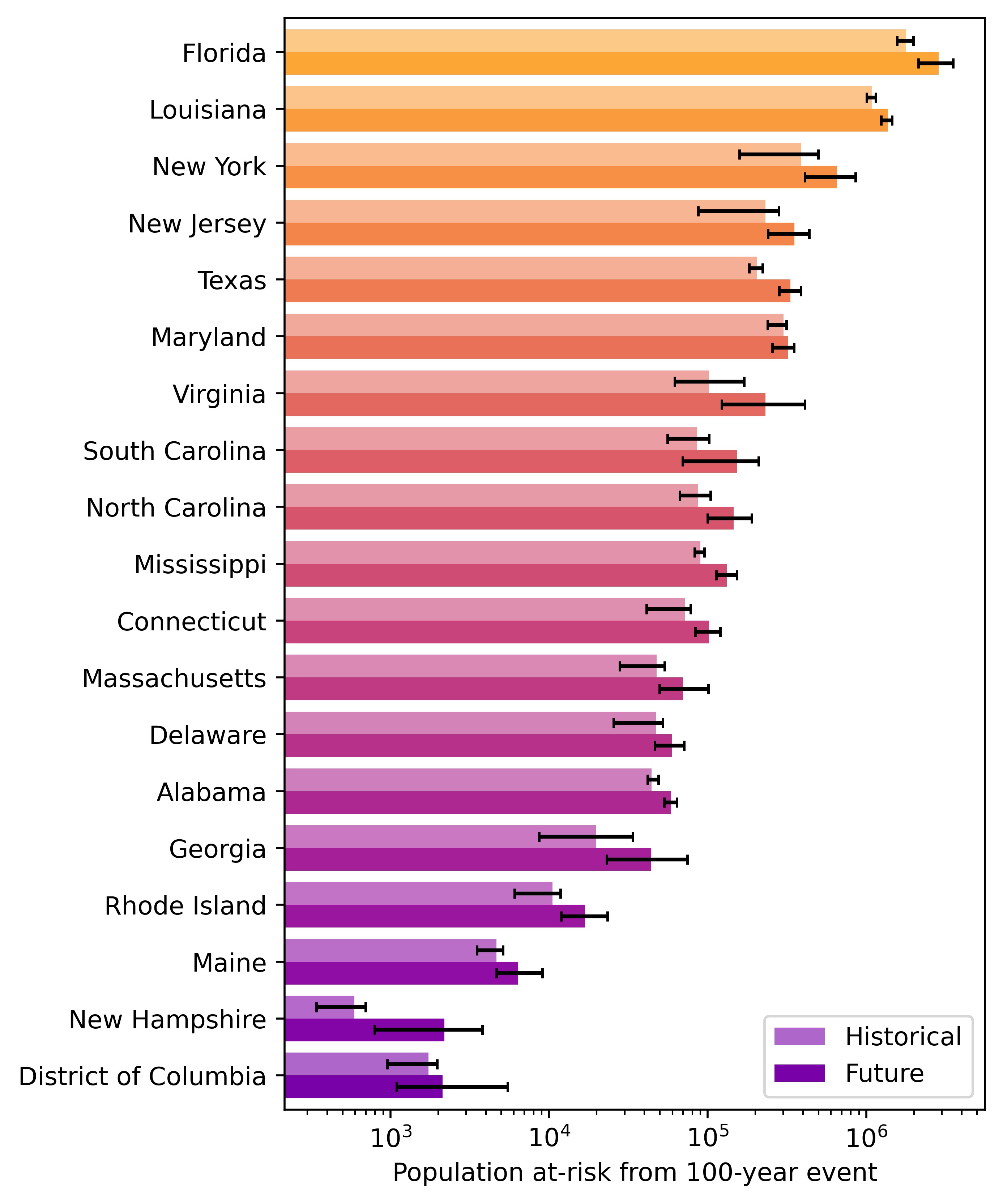}
        \caption{Log scaling}
        \label{fig:inundation-uncertainty-logscaled}
    \end{subfigure}

    \caption{Median estimate of population at risk from 100-year flood event in the historical and future periods for each state, along with 90\% uncertainty intervals, (\textbf{a}) linearly scaled, and (\textbf{b}) logarithmically scaled for clarity in less-affected states.}
    \label{fig:inundation-uncertainty}
\end{figure}

% State eCDFs
\begin{figure}
    \centering

    \includegraphics[width=\textwidth]{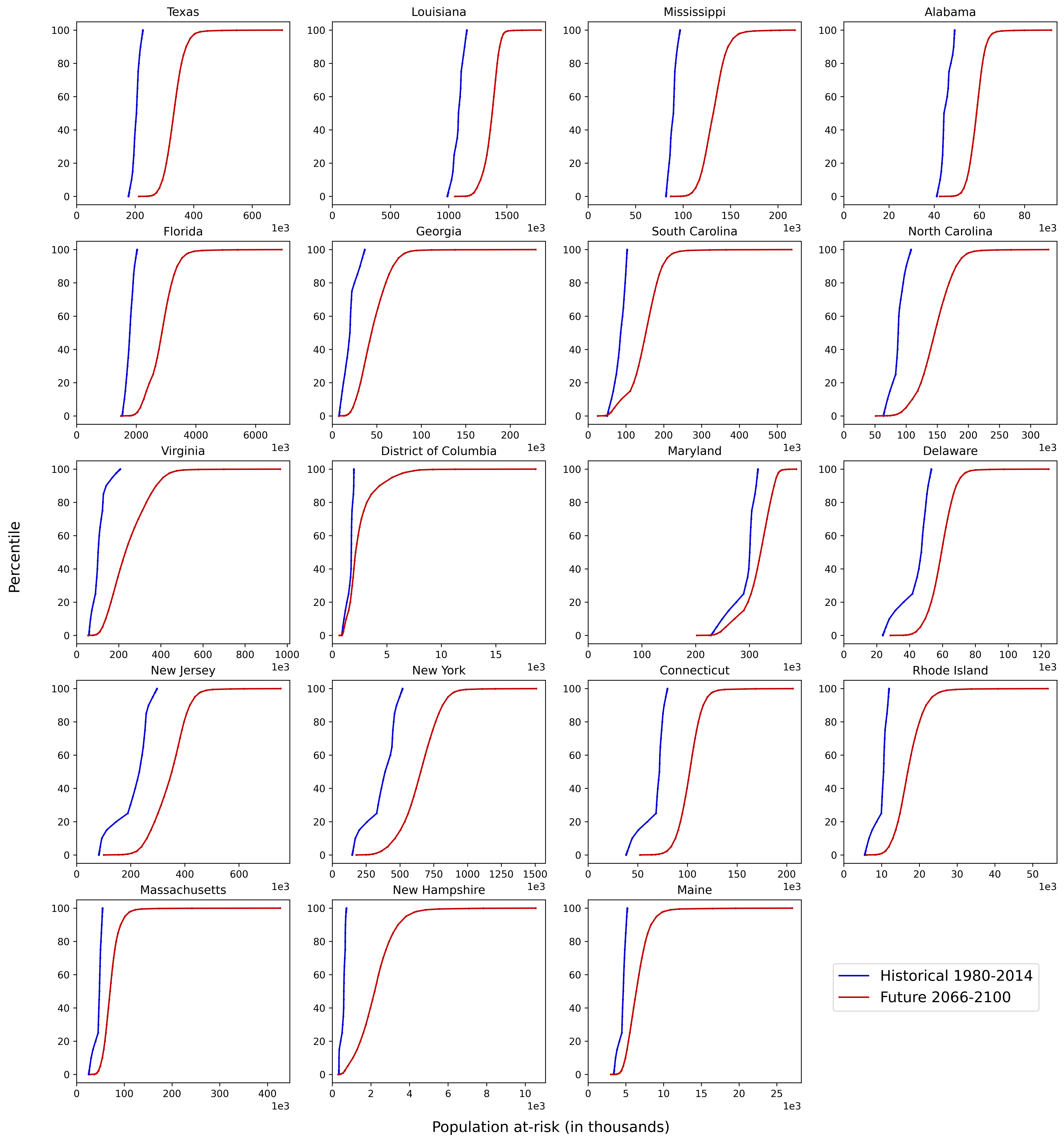}
    \caption{Empirical cumulative distribution functions (eCDFs) of population at risk from DeepSurge-modeled 100-year storm surge flooding for each state.}
    \label{fig:state-percentiles}
\end{figure}

% ADCIRC domain
\begin{figure}
    \centering
    \begin{subfigure}[]{0.4\textwidth}
        \centering
        \includegraphics[width=\textwidth]{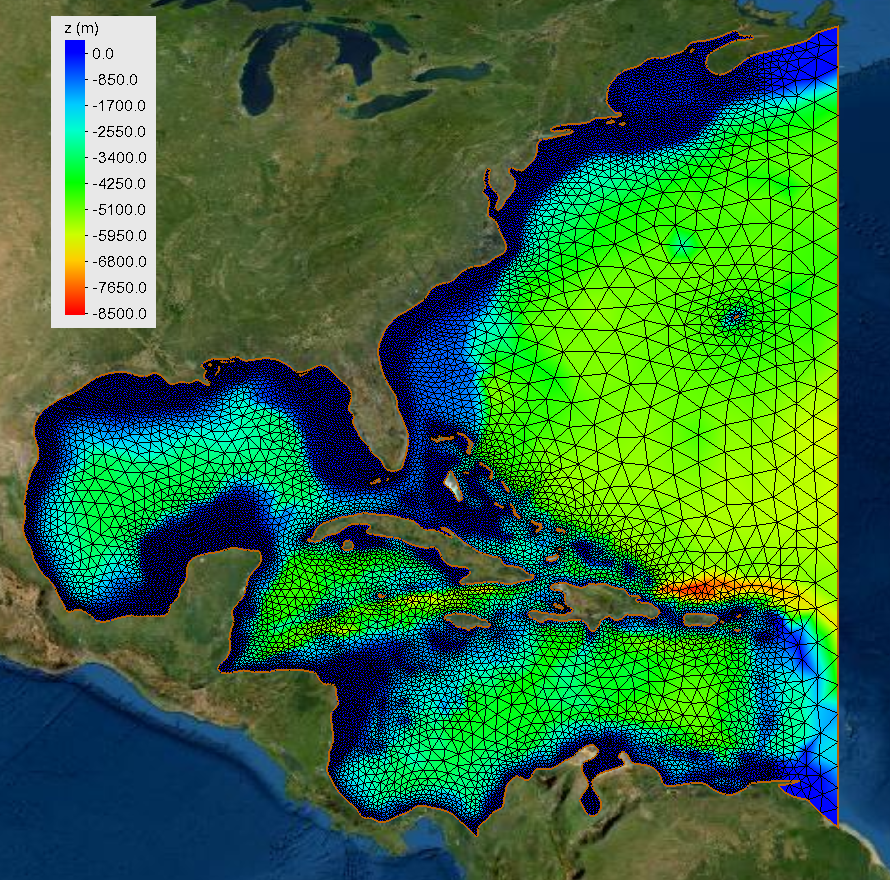}
        % \caption{\textcolor{red}{Supplement.} ADCIRC domain, triangular elements of unstructured grid, and bathymetry colormap.}
        \label{fig:adcirc-bathy-grid}
    \end{subfigure}
    \begin{subfigure}[]{0.55\textwidth}
        \includegraphics[width=\textwidth]{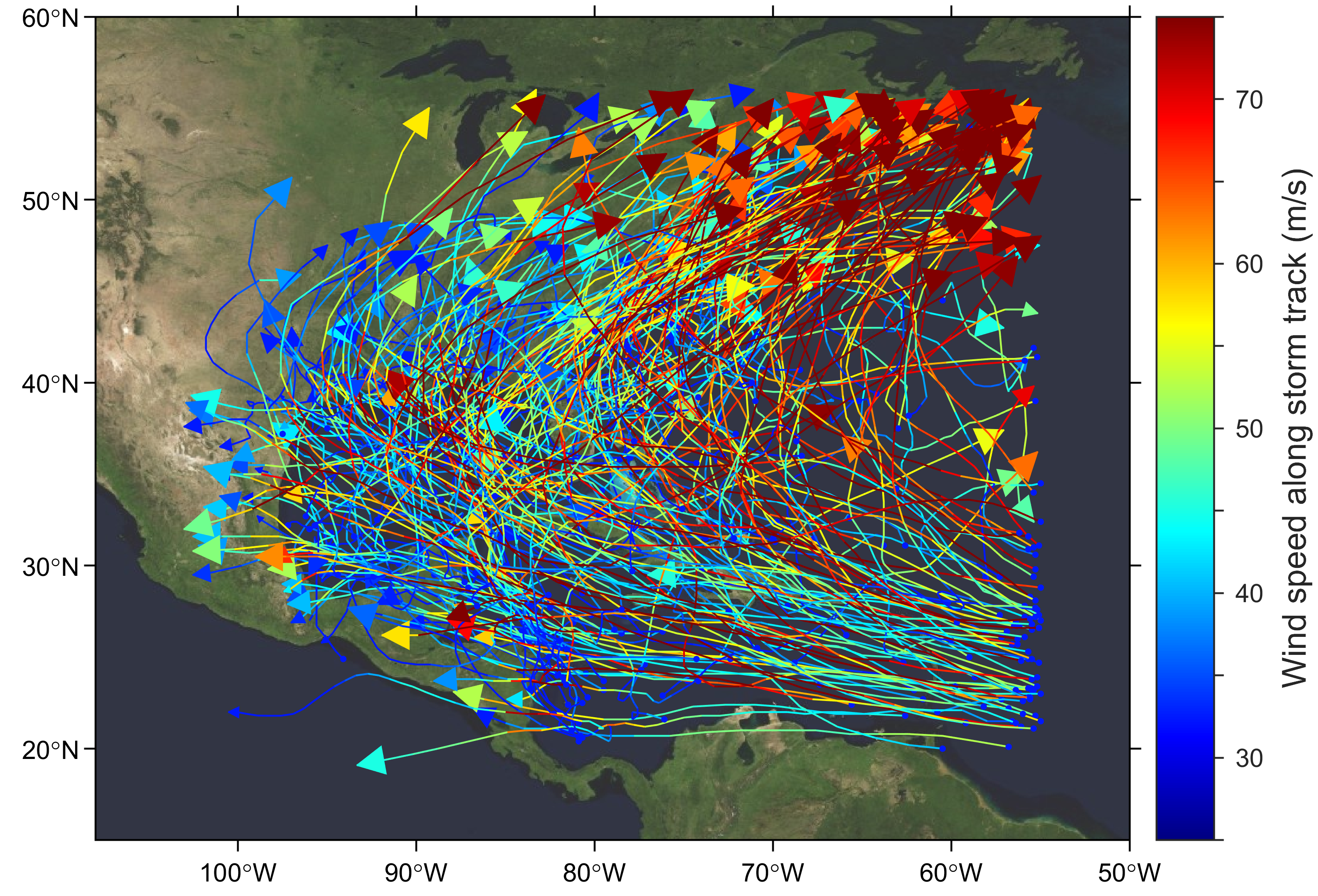}
        % \caption{\textcolor{red}{Supplement.} Historical TC tracks simulated by ADCIRC.}
        \label{fig:adcirc-hurricane-tracks}
    \end{subfigure}
    \caption{(a) ADCIRC domain, triangular elements of unstructured grid, and bathymetry colormap. (b) Historical TC tracks simulated by ADCIRC used for training.}
    \label{fig:adcirc-grid-and-storms}
\end{figure}

% DeepSurge tide gauge bias
\begin{figure}
    \centering
    \includegraphics[width=\textwidth]{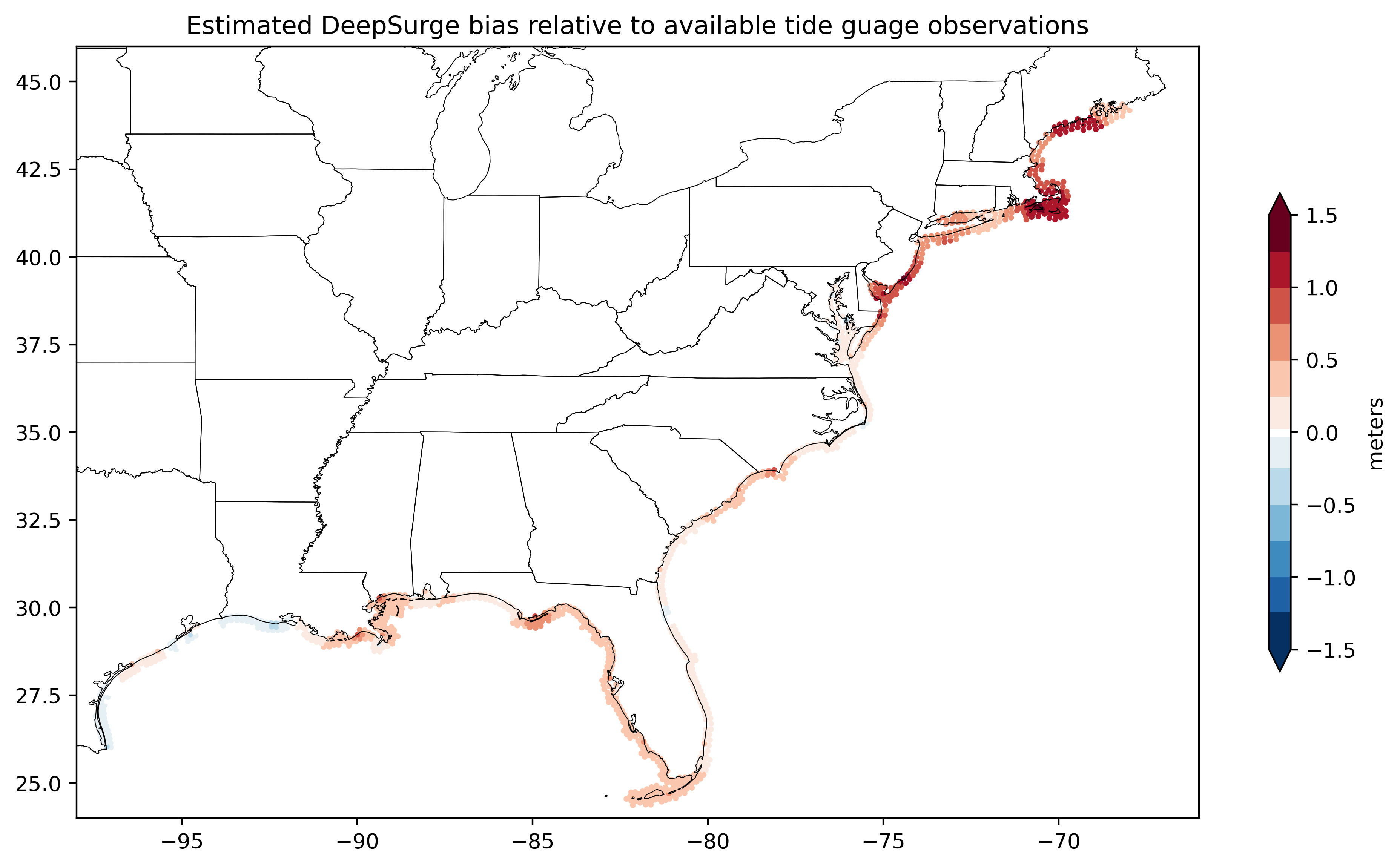}
    \caption{Estimated spatial mean bias of DeepSurge relative to tide gauge observations. Note that small sample sizes make the error estimates in the Northeast much less robust.}
    \label{fig:deepsurge-bias}
\end{figure}

% Needham eCDFs
\begin{figure}
    \centering
    \includegraphics[width=0.8\linewidth]{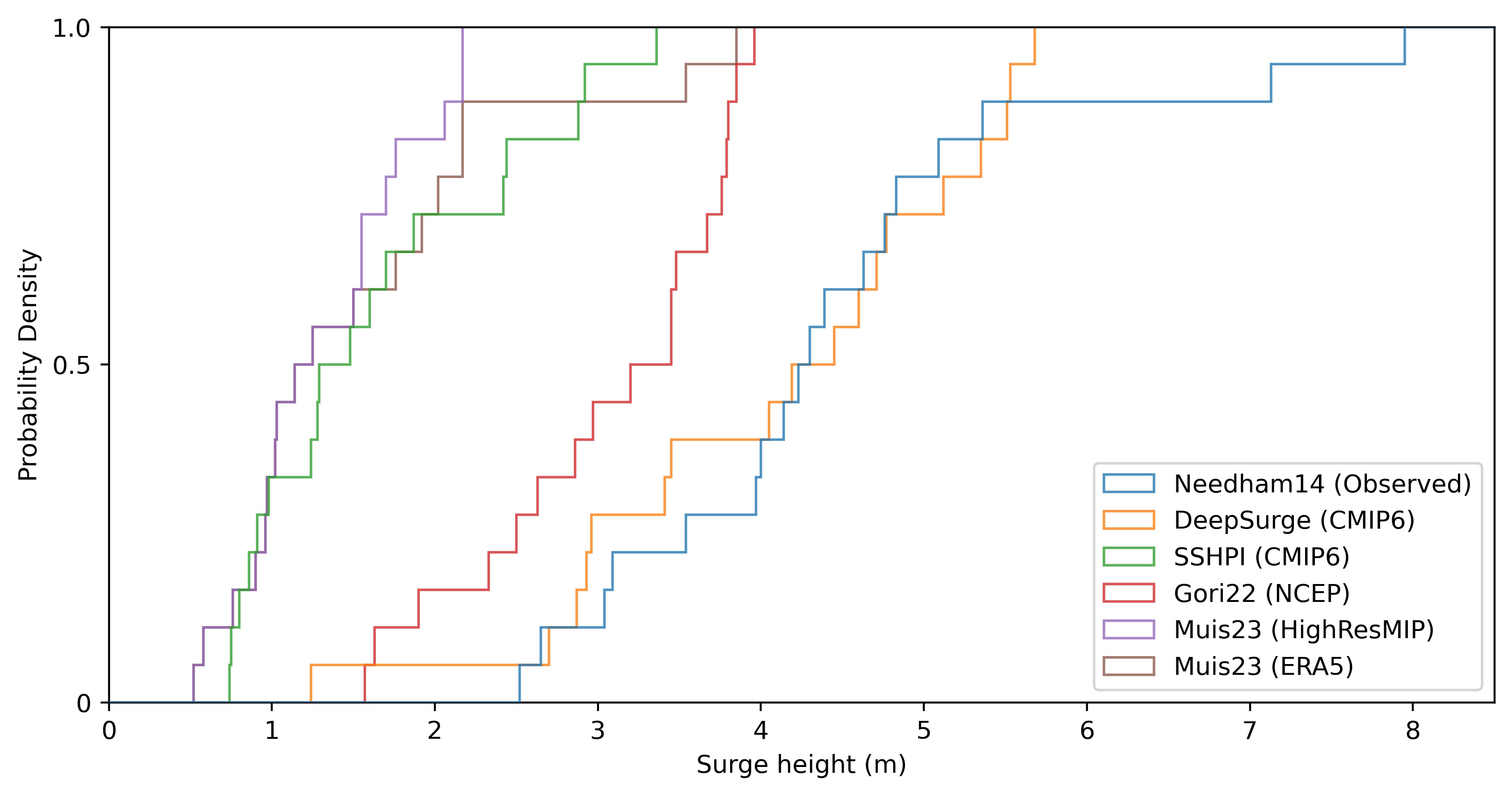}
    \caption{Empirical cumulative distribution functions (eCDFs) of historical 100-year surge heights from 18 locations across the Gulf Coast estimated by the \citet{needham_data-driven_2014} compared to each modeling method. The methods are: DeepSurge (ours), \citet{gori_tropical_2022}, \citet{muis_global_2023}, and SSHPI \citep{islam_new_2021}, with the climate data used to force the model listed in parentheses.}
    \label{fig:needham-cdfs}
\end{figure}

% Method correlation matrix
\begin{figure}
    \centering
    \begin{subfigure}[]{0.49\textwidth}
        \centering
        \includegraphics[width=\textwidth]{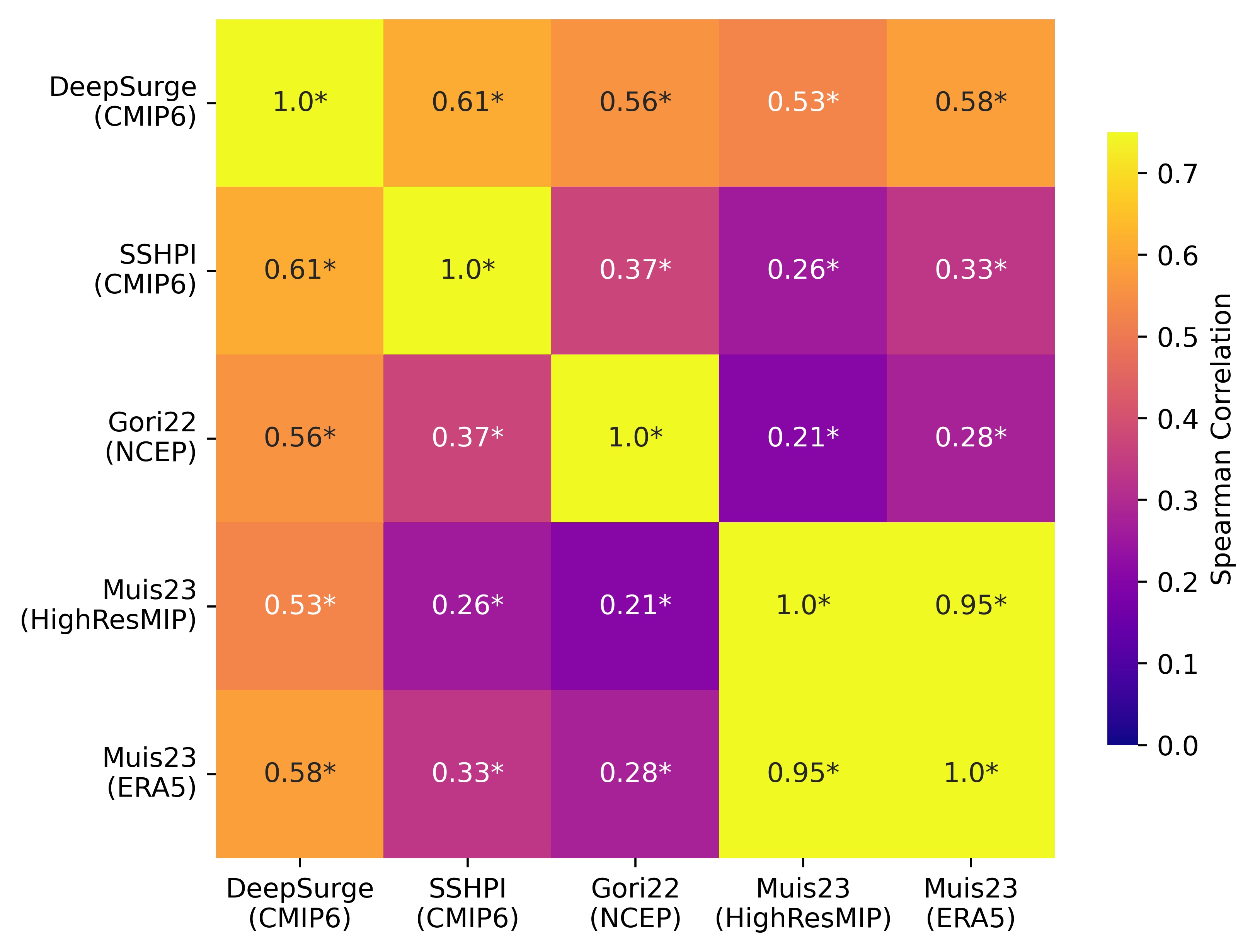}
        \caption{Historical period}
        \label{fig:modelcomp-hist-corr}
    \end{subfigure}
    \hfill
    \begin{subfigure}[]{0.49\textwidth}
        \centering
        \includegraphics[width=\textwidth]{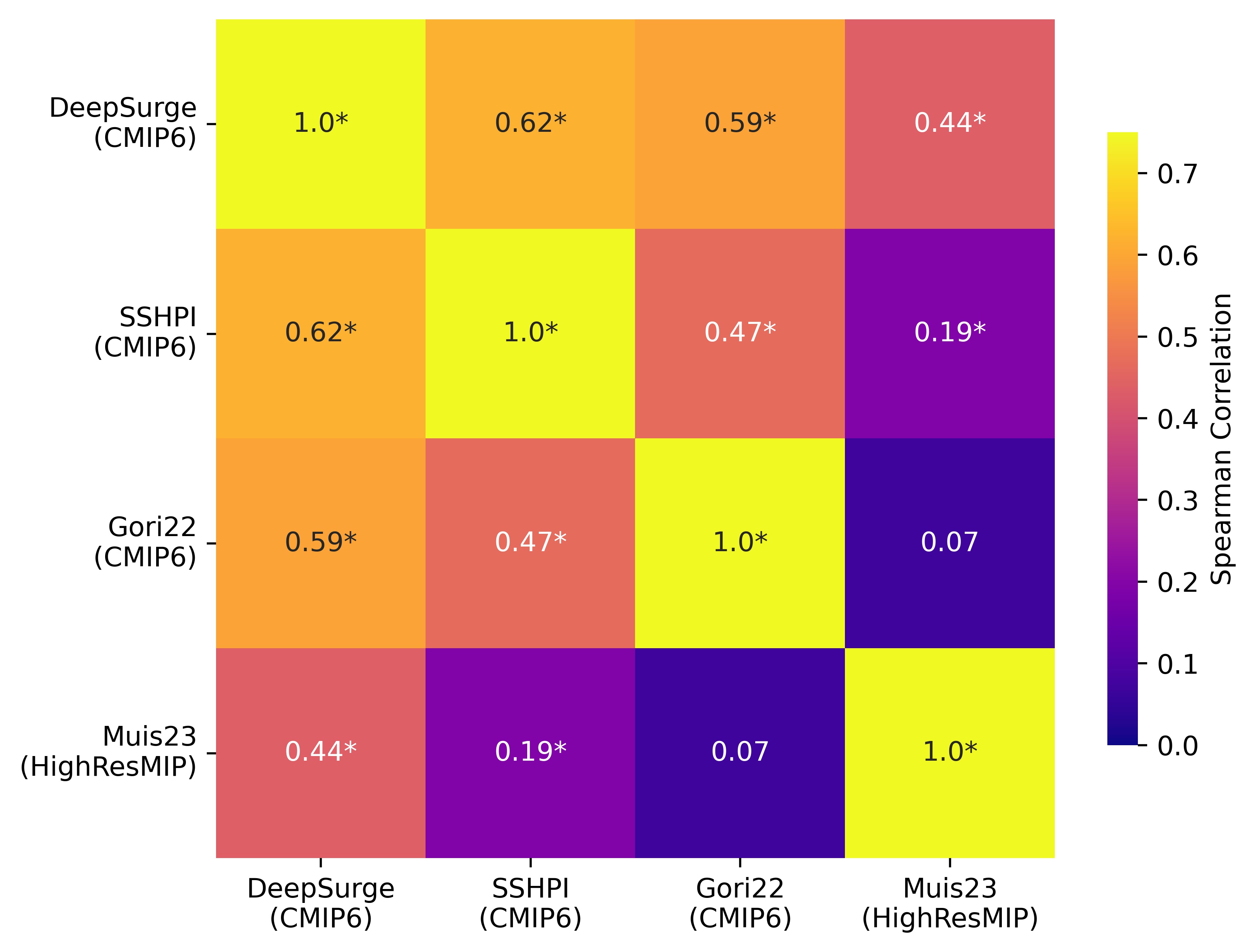}
        \caption{Future period}
        \label{fig:modelcomp-futu-corr}
    \end{subfigure}
    \hfill
    \caption{Spearman correlations between methods for the spatial distribution of the 100-year return level, as estimated in each method's respective (\textbf{a}) historical and (\textbf{b}) future period. Asterisks indicate significance at the 95\% level. Note that these methods have varying definitions of the historical and future time period. The methods are: DeepSurge (ours), \citet{gori_tropical_2022}, \citet{muis_global_2023}, and SSHPI \citep{islam_new_2021}, with the climate data used to force the model listed in parentheses.}
    \label{fig:modelcomp}
\end{figure}

% Surge comparison maps
\begin{figure}
    \centering
    \begin{subfigure}[]{0.49\textwidth}
        \centering
        \includegraphics[width=\textwidth]{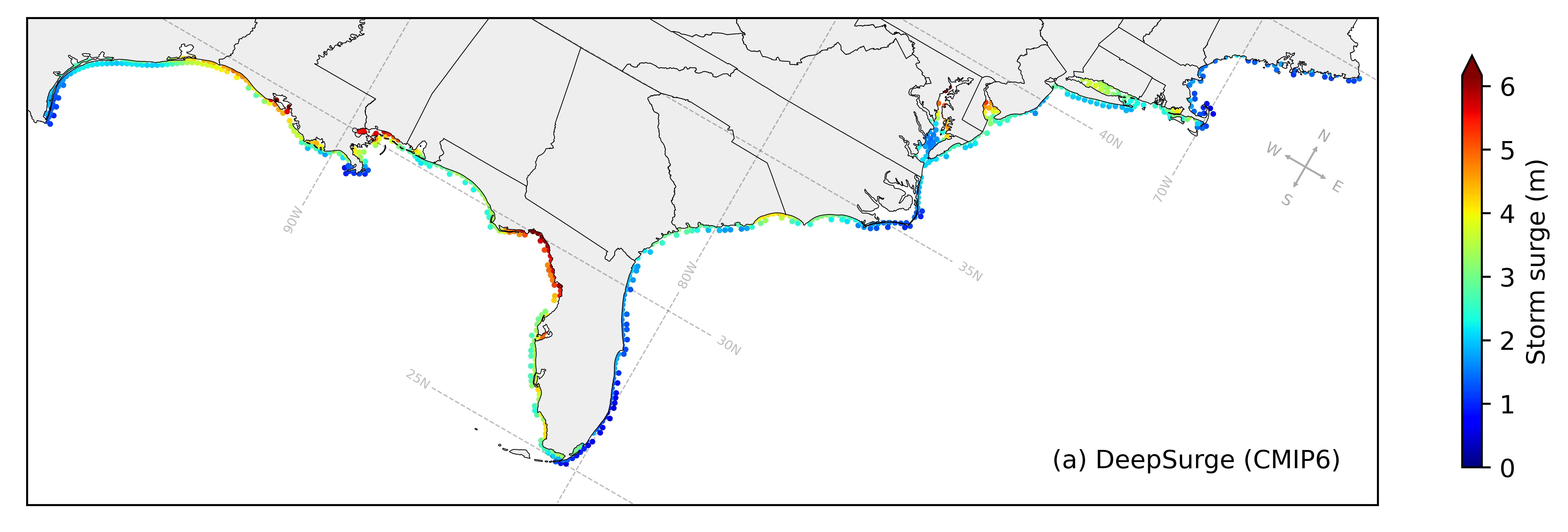}
        \caption{DeepSurge (RAFT-CMIP6 1980-2014 forcing)}
        % \label{fig:hist-storm-direction}
    \end{subfigure}
    \hfill
    \begin{subfigure}[]{0.49\textwidth}
        \centering
        \includegraphics[width=\textwidth]{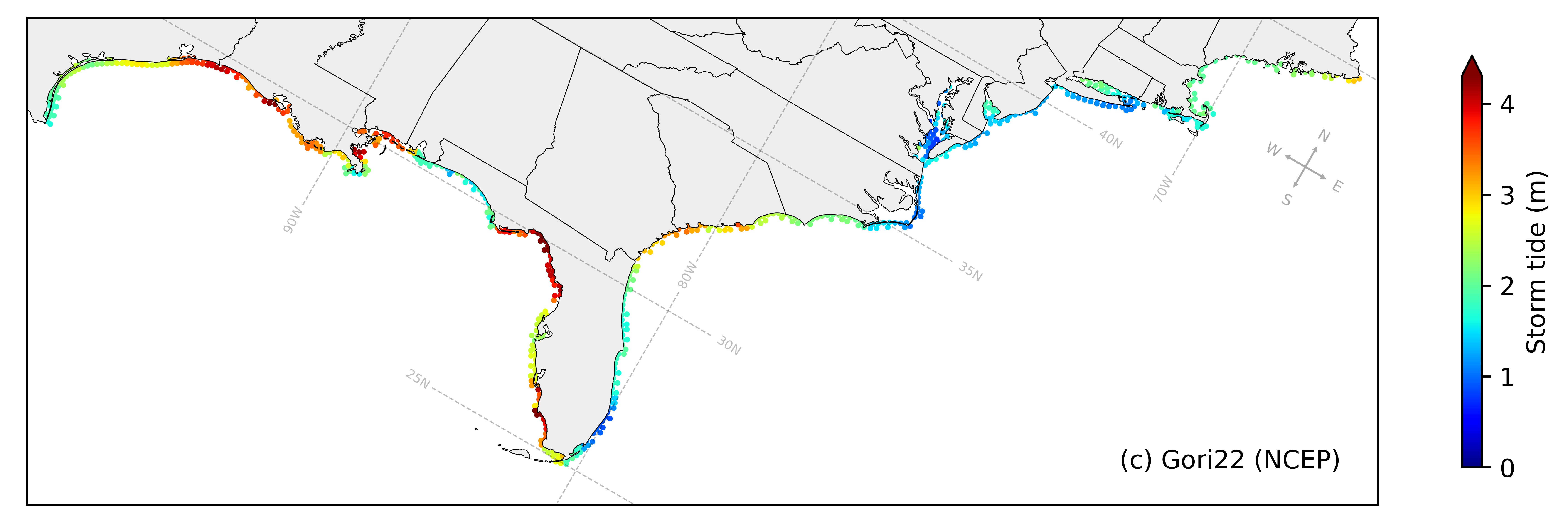}
        \caption{\citet{gori_tropical_2022} (NCEP 1980-2005 forcing). Note that these estimates are in terms of storm tide, not storm surge.}
        % \label{fig:change-storm-direction}
    \end{subfigure}

    \begin{subfigure}[]{0.49\textwidth}
        \centering
        \includegraphics[width=\textwidth]{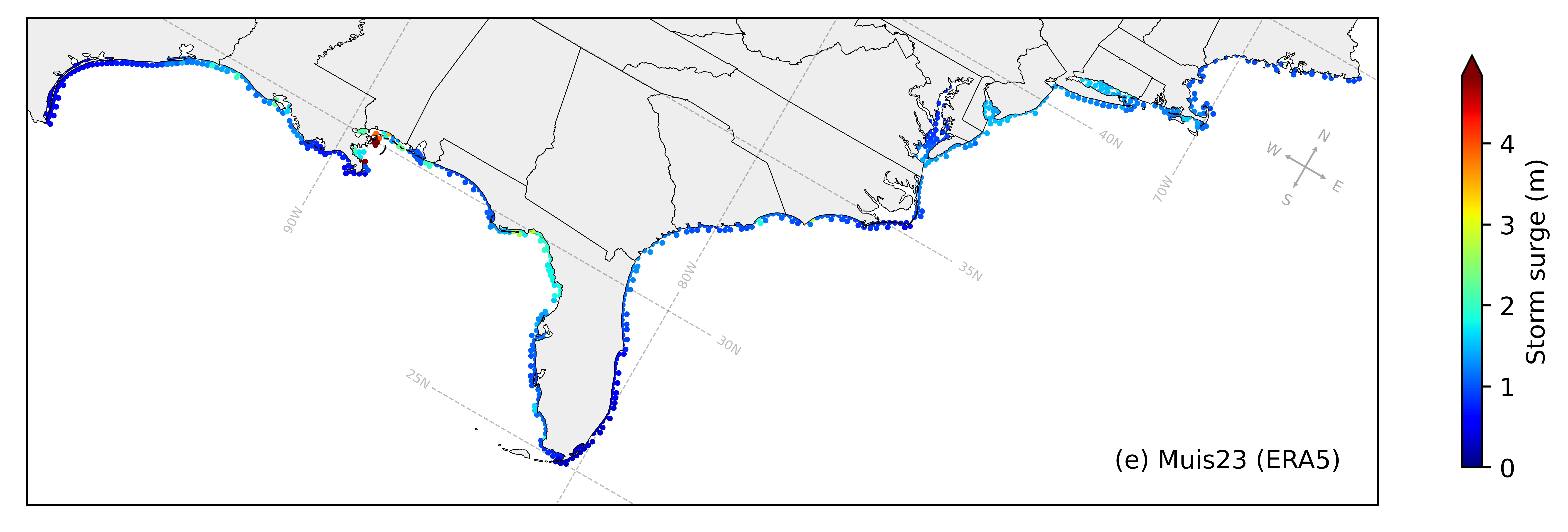}
        \caption{\citet{muis_global_2023} (ERA5 1985-2014 forcing)}
        % \label{fig:hist-storm-direction}
    \end{subfigure}
    \hfill
    \begin{subfigure}[]{0.49\textwidth}
        \centering
        \includegraphics[width=\textwidth]{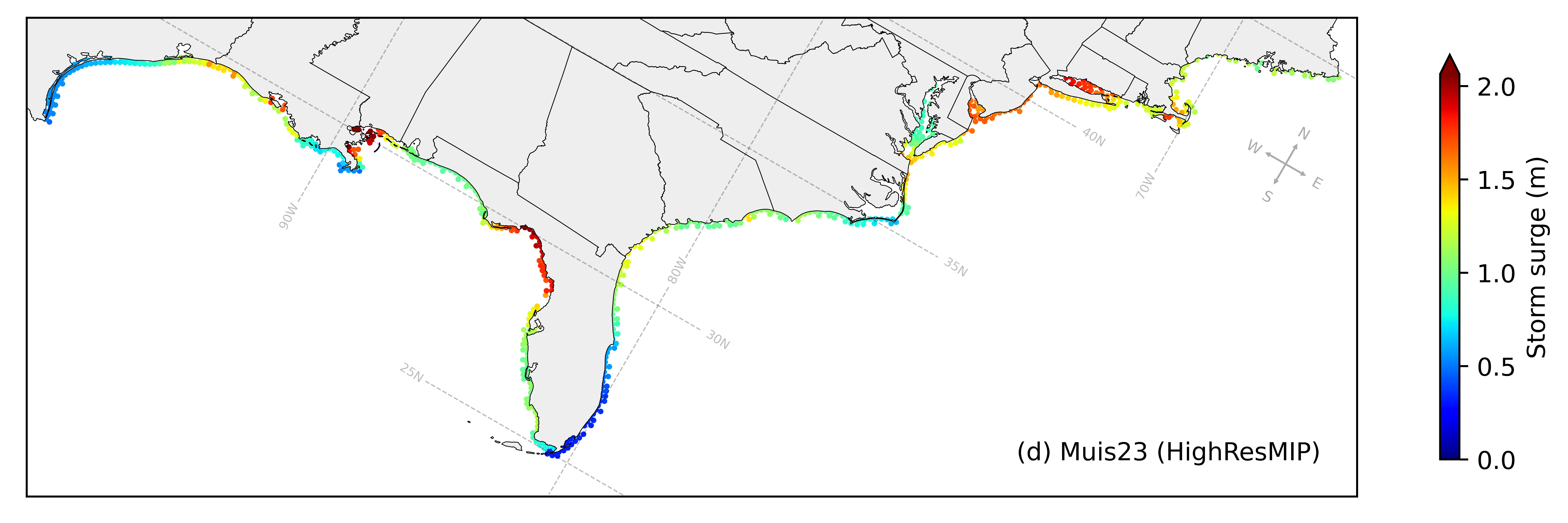}
        \caption{\citet{muis_global_2023} (HighResMIP 1950-2014 forcing)}
        % \label{fig:change-storm-direction}
    \end{subfigure}

    \begin{subfigure}[]{0.49\textwidth}
        \centering
        \includegraphics[width=\textwidth]{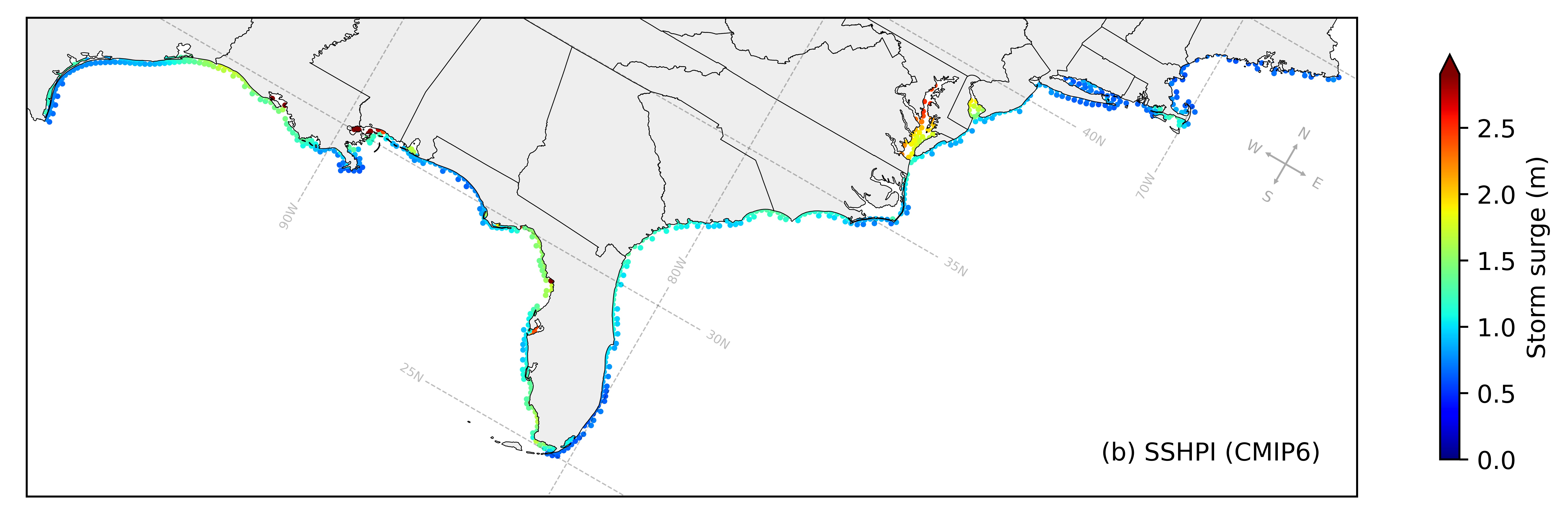}
        \caption{SSHPI (RAFT-CMIP6 1980-2014 forcing)}
        % \label{fig:hist-storm-direction}
    \end{subfigure}
    \caption{The 100-year surge level estimates from five modeling methods, for their respective historical periods. Note that the vastly varying magnitudes necessitate different colorscales, with the colorscale maximum set to the 99th percentile of the data for each.}
    \label{fig:modelcomp-100yr-events}
\end{figure}

% FEMA HWM
\begin{figure}
    \centering
    \begin{subfigure}[]{0.8\textwidth}
        \centering
        \includegraphics[width=\textwidth]{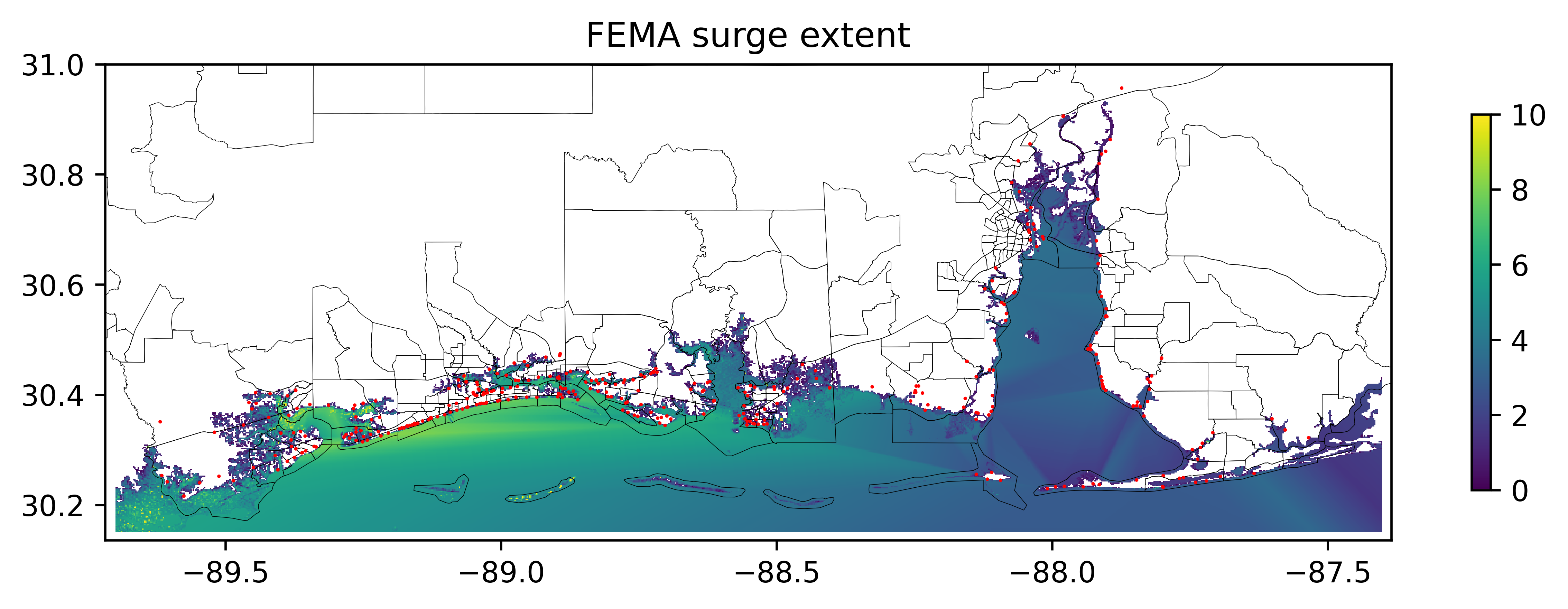}
        \caption{Inundation surface estimated from FEMA high-water marks. Red dots indicate location of high water marks.}
        \label{fig:fema-hwm-surge}
    \end{subfigure}
    \begin{subfigure}[]{0.8\textwidth}
        \centering
        \includegraphics[width=\textwidth]{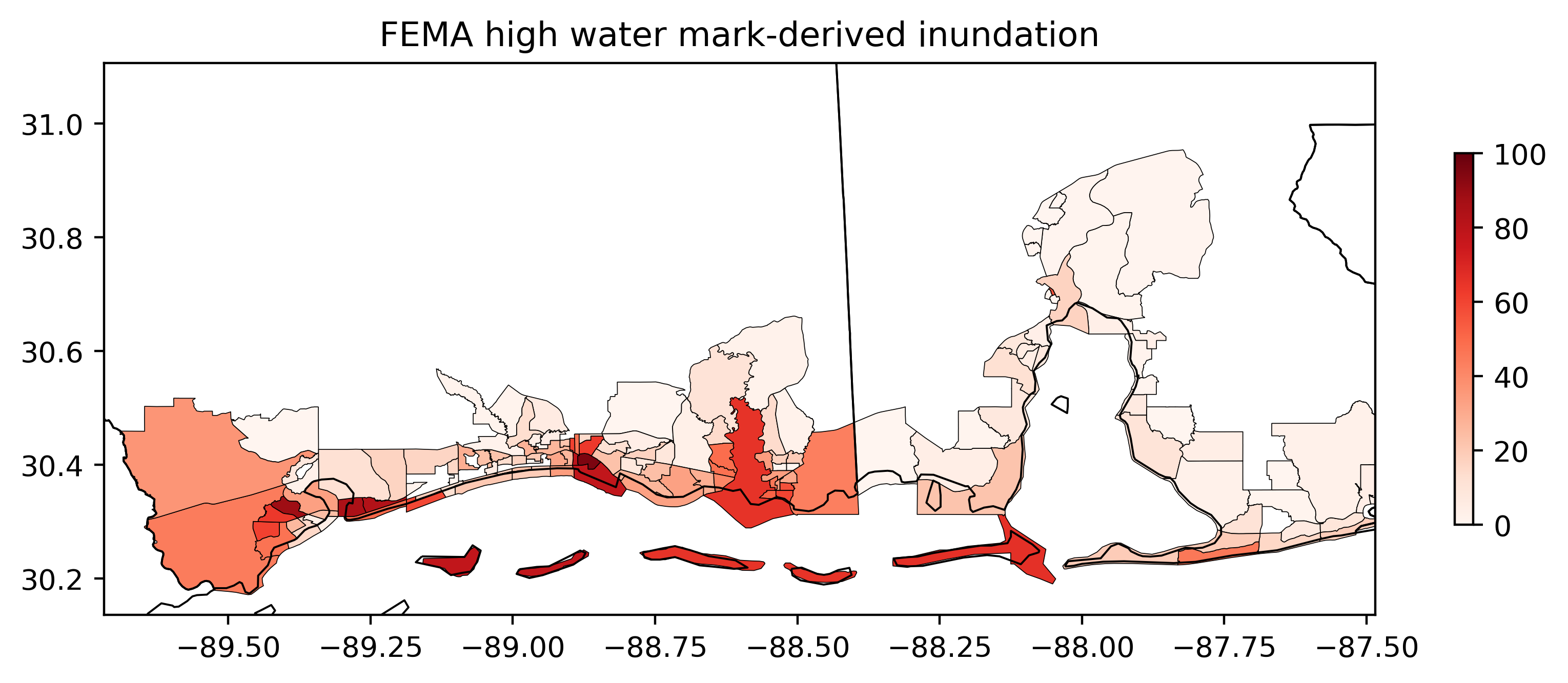}
        \caption{Population affected estimated from FEMA high-water marks}
        \label{fig:fema-hwm-inundation}
    \end{subfigure}
    \begin{subfigure}[]{0.8\textwidth}
        \centering
        \includegraphics[width=\textwidth]{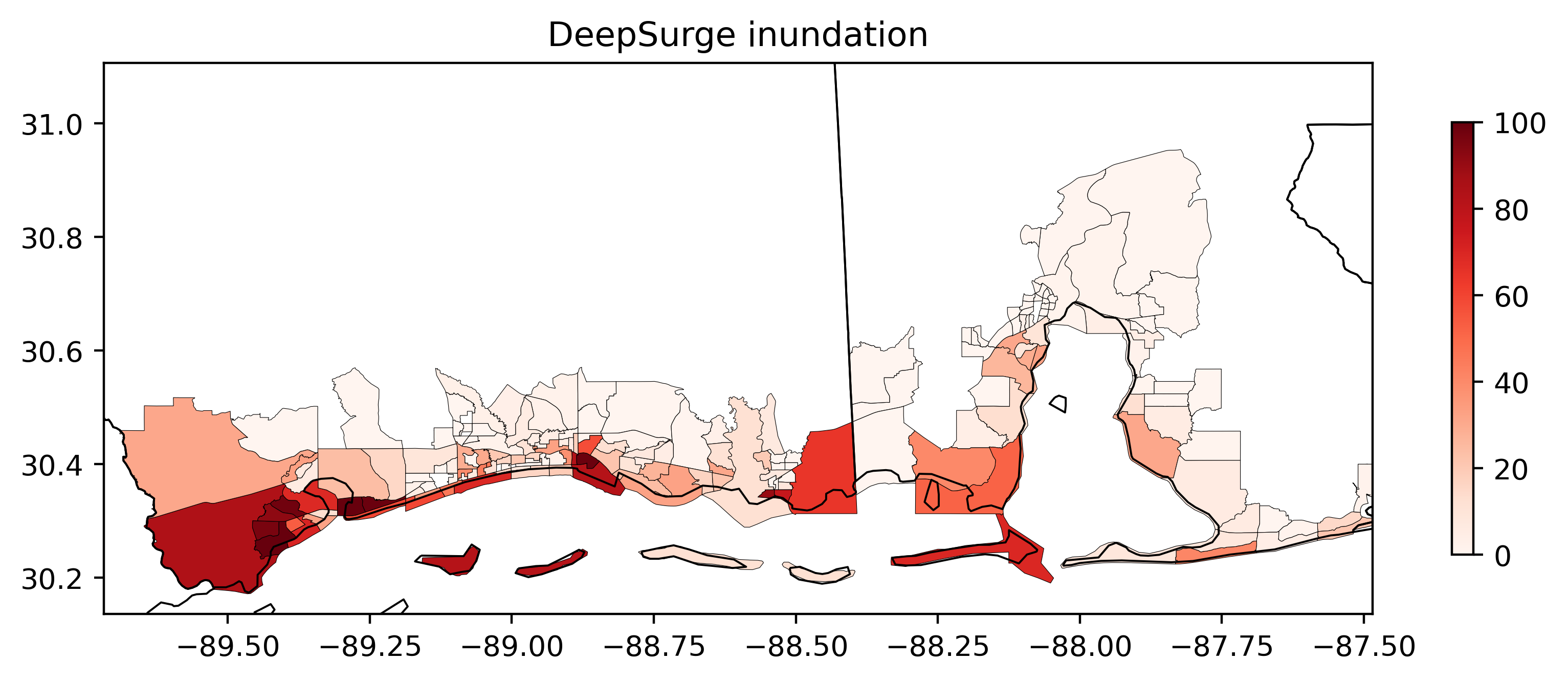}
        \caption{Population affected estimated by DeepSurge and CA-Surge}
        \label{fig:deepsurge-hwm-inundation}
    \end{subfigure}
    \caption{Inundated (\textbf{a}) area and (\textbf{b}) population at the census-tract level estimated from FEMA high-water marks from Hurricane Katrina (2005), in Mississipi (left half) and Alabama (right half). (\textbf{c}) Inundated population estimated by our method. Note that since census tracts have roughly the same number of residents, large tracts are not any more important that small tracts. Some islands are part of mainland census tracts.}
    \label{fig:hwm-comp}
\end{figure}

% Crowell
\begin{figure}
    \centering
    \includegraphics[width=0.8\textwidth]{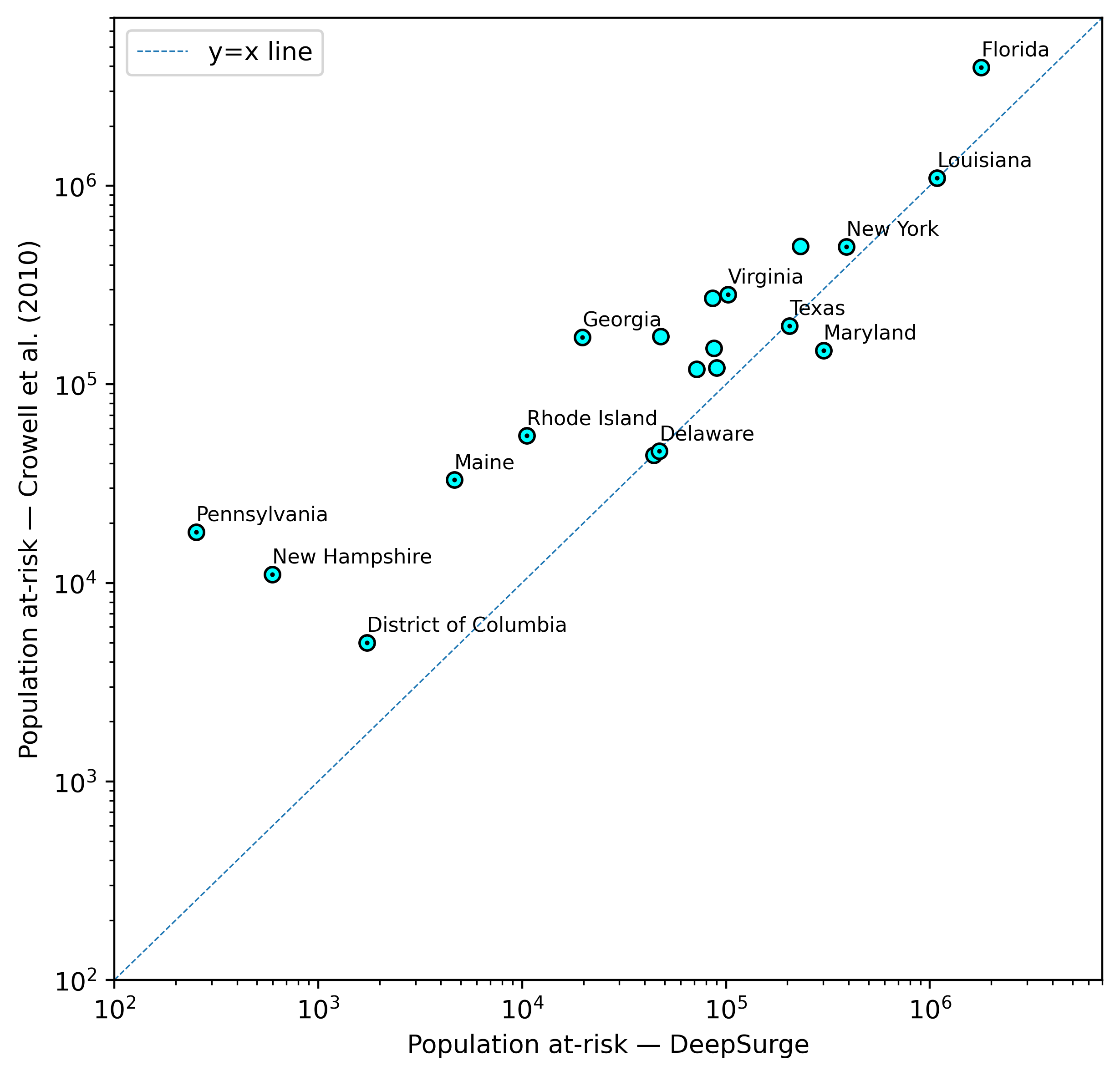}
    \caption{Comparison between historical 100-year inundation estimates from \citet{crowell_estimate_2010} and our method. They exhibit Pearson and Spearman correlations of 0.95 and 0.87 respectively ($p \ll 0.01$).}
    \label{fig:crowell-comp}
\end{figure}

% SSHPI drivers of change
\begin{figure}
    % \centering
    \begin{subfigure}[]{0.42\textwidth}
        \centering
        \includegraphics[width=\textwidth]{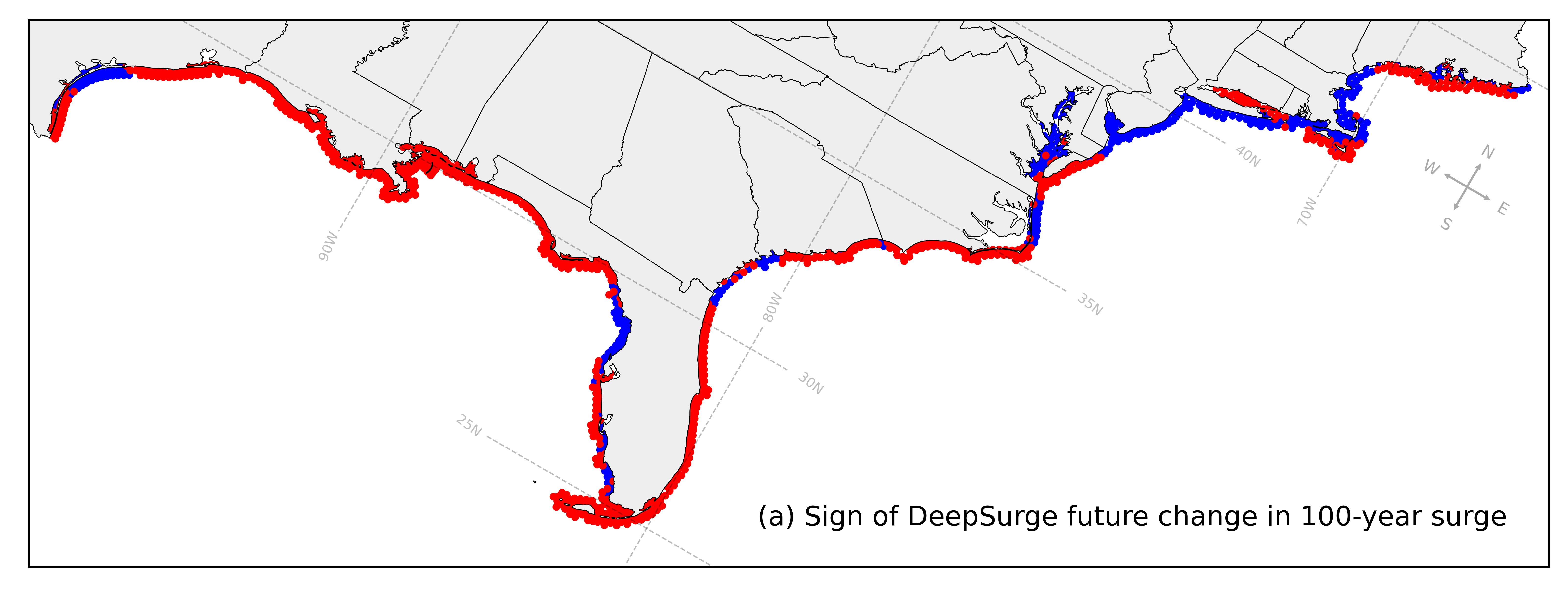}
        % \caption{DeepSurge-projected sign of future change (without sea-level rise). Red is positive, blue is negative.}
        \phantomsubcaption
        \label{fig:deepsurge-sign}
    \end{subfigure}
    \hspace{3em}
    \begin{subfigure}[]{0.42\textwidth}
        % \centering
        \includegraphics[width=\textwidth]{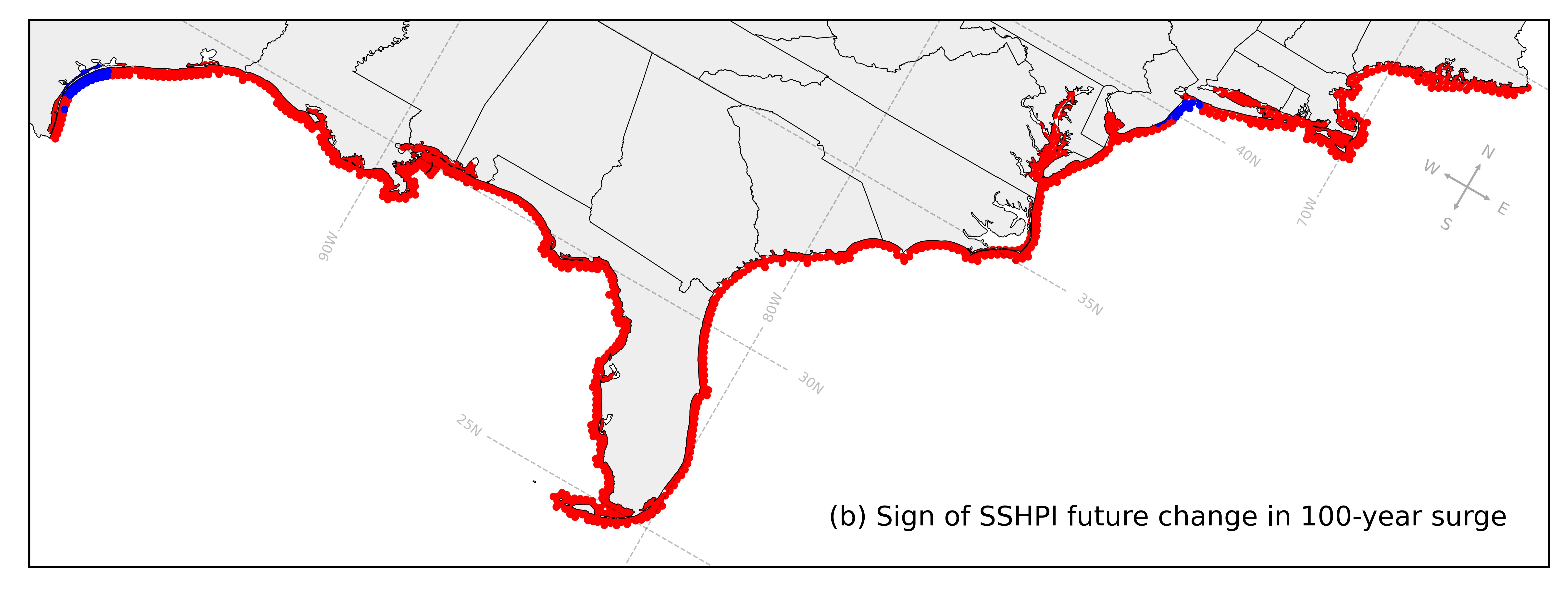}
        % \caption{SSHPI-projected sign of future change (without sea-level rise). Red is positive, blue is negative.}
        \phantomsubcaption
        \label{fig:sshpi-sign}
    \end{subfigure}

    \begin{subfigure}[]{0.49\textwidth}
        \centering
        \includegraphics[width=\textwidth]{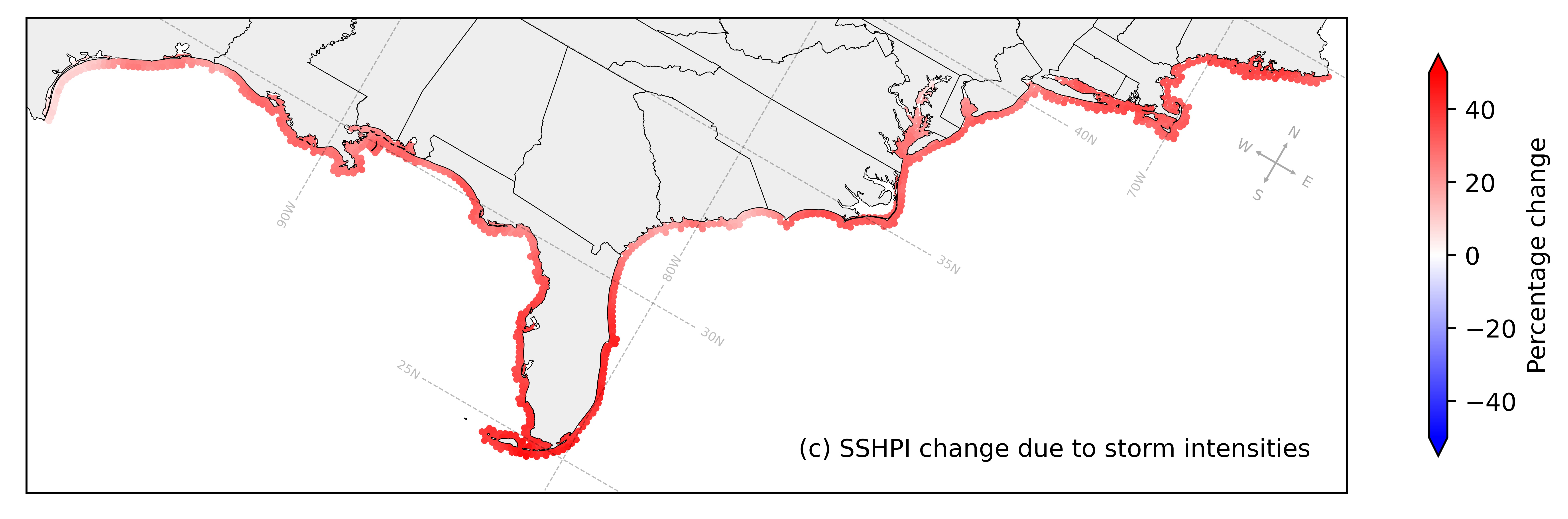}
        % \caption{Storm intensity term}
        \phantomsubcaption
        \label{fig:sshpi-vterm}
    \end{subfigure}
    \hfill
    \begin{subfigure}[]{0.49\textwidth}
        \centering
        \includegraphics[width=\textwidth]{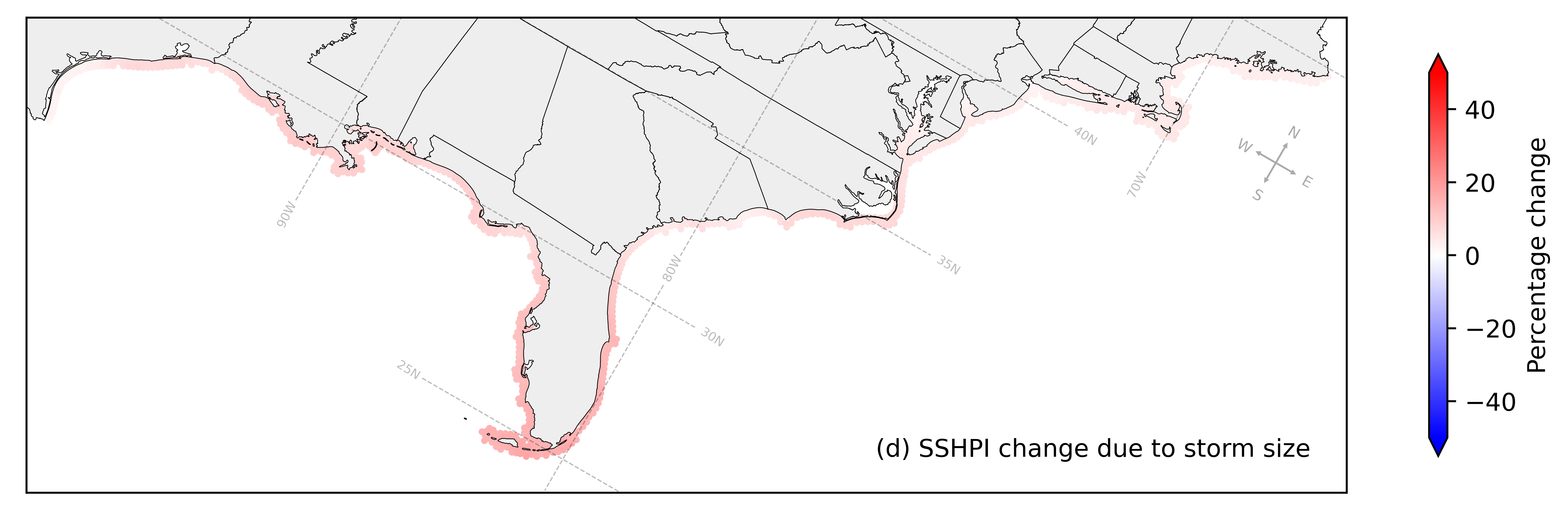}
        % \caption{Storm size term}
        \phantomsubcaption
        \label{fig:sshpi-rterm}
    \end{subfigure}
    % \hfill
    
    \begin{subfigure}[]{0.49\textwidth}
        \centering
        \includegraphics[width=\textwidth]{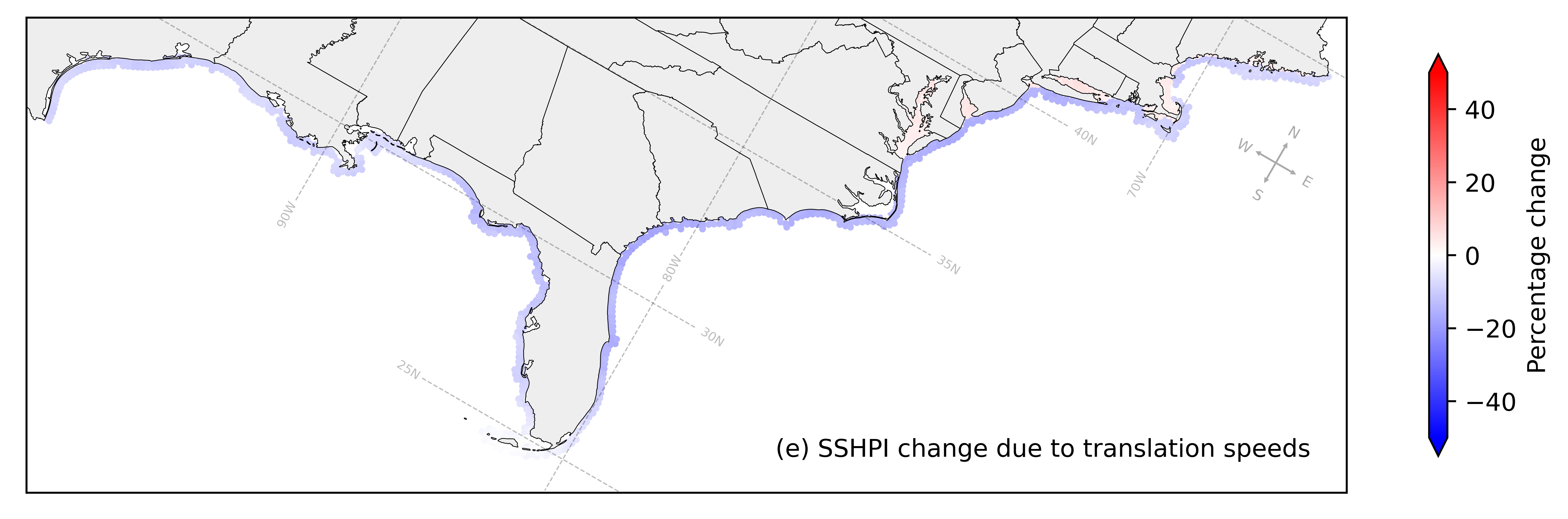}
        % \caption{Translation speed term}
        \phantomsubcaption
        \label{fig:sshpi-sterm}
    \end{subfigure}
    \hspace{0.4em}
    \begin{subfigure}[]{0.42\textwidth}
        \centering
        \includegraphics[width=\textwidth]{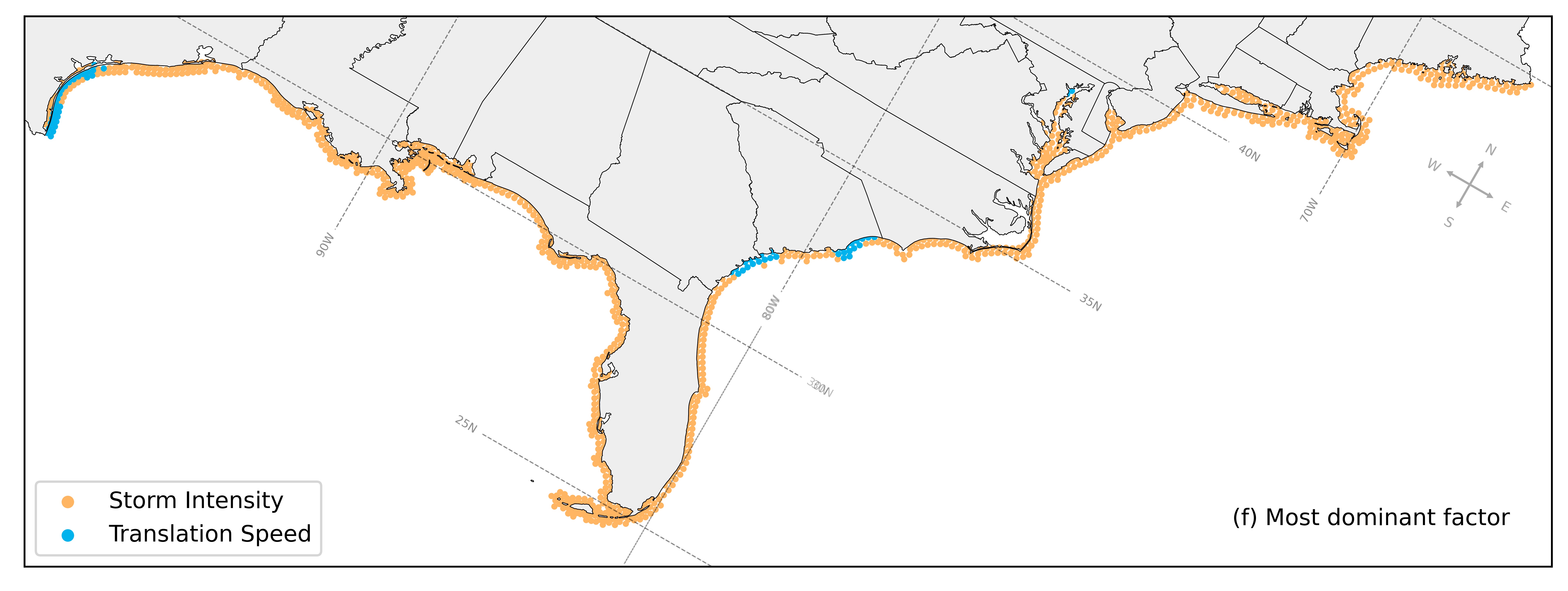}
        % \caption{Most dominant SSHPI factor}
        \phantomsubcaption
        \label{fig:sshpi-dominant}
    \end{subfigure}

    \caption{Sign of future change in 100-year surge height, as projected by (\textbf{a}) DeepSurge, and (\textbf{b}) SSHPI (red is positive, blue is negative). Percentage change in the future-climate SSHPI-derived surge for all events with return periods of $\ge100$ years, due to (\textbf{c}) storm intensity (\textbf{d}) storm size and (\textbf{e}) storm translation speed. (\textbf{f}) The largest contributing factor at each node.}
    \label{fig:sshpi-factors}
\end{figure}

\clearpage

\bibliographystyle{icml2023}
\bibliography{Surge}